\documentclass[fleqn,usenatbib]{mnras}

\usepackage{newtxtext,newtxmath}

\usepackage[T1]{fontenc}
\usepackage[usenames]{color}
\usepackage{colortbl}
\usepackage{float}

\DeclareRobustCommand{\VAN}[3]{#2}
\let\VANthebibliography\thebibliography
\def\thebibliography{\DeclareRobustCommand{\VAN}[3]{##3}\VANthebibliography}

\usepackage{graphicx}	
\usepackage{amsmath}	

\title[The gMOSS: galaxy survey]{The gMOSS: the galaxy survey and galaxy populations of the large homogeneous field}

\author[Grokhovskaya A., Dodonov S. N., Movsessian T.A., Kotov S.S.]{
Grokhovskaya A.,$^{1,2}$\thanks{E-mail: grohovskaya.a@gmail.com}
 Dodonov S.N.,$^{1,2}$
 Movsessian T.A.,$^{3}$
 Kotov S.S.$^{1,2}$
\\
$^{1}$ Special Astrophysical Observatory, Russian Academy of Sciences, Nizhny Arkhyz 369167, Russia\\
$^{2}$ Institute of Applied Astronomy of the Russian Academy of Sciences, Kutuzov Quay 10, St. Petersburg, 191187, Russia\\
$^{3}$ Byurakan Astrophysical Observatory, 0213 Aragatzotn prov., Armenia
}

\date{Accepted XXX. Received YYY; in original form ZZZ}

\pubyear{2021}

\begin{document}
\label{firstpage}
\pagerange{\pageref{firstpage}--\pageref{lastpage}}
\maketitle

\begin{abstract}
We present the gMOSS (Galaxies of Medium-band One-meter Schmidt telescope Survey) catalog of $\sim$ 19,000 galaxies in 20 filters (4 broadband SDSS and 16 medium-band filters). We observed 2.386 $\mathrm{deg^2}$ on the central part of the HS47.5-22 field with the 1-m Schmidt telescope of the Byurakan Astrophysical Observatory. The gMOSS is a complete flux-limited sample of galaxies with a threshold magnitude of $r$ SDSS $\le$ 22.5 AB. From photometric measurements with 16 medium-band filters and $u$ SDSS, we get spectral energy distributions for each object in the field, which are used for further analysis. Galaxy classification and photometric redshift estimation based on spectral template matching with \textsc{zebra} software. The obtained redshift accuracy is $\sigma_\mathrm{{NMAD}} < 0.0043$. Using the SED-fitting \textsc{cigale} code, we obtained the main properties of the stellar population of galaxies, such as rest-frame $(u - r)_{\mathrm{res}}$ colour, stellar mass, extinction, and mass-weighted age with a precision of $0.16 \pm 0.07$ mag, $0.14 \pm 0.04$ dex, $0.27 \pm 0.1$ mag, and $0.08 \pm 0.04$ dex, respectively. Using a  dust-corrected colour-mass diagram, we divided the full sample into populations of red and blue galaxies and considered the dependencies between stellar mass and age. Throughout cosmic time, red sequence galaxies remain older and more massive than blue cloud galaxies. The star formation history of a complete subsample of galaxies selected in the redshift range $0.05\le z\le0.015$ with <$\mathrm{log} M \mathrm{>}_\mathrm{[M_\odot]}$>8.3 shows an increase in the SFRD up to $z\sim3$, under the results obtained in earlier studies.

\end{abstract}




\section{Introduction}

The evolution and physical properties of galaxies require statistical studies with numerous objects. Using spectroscopic redshifts is most preferable for analyzing the evolution of the physical properties of galaxies with the redshift. Spectroscopic redshifts were widely used in research on relatively bright galaxies with small redshift (e.g. in \citealt*{Peng2010}). However, for samples of tens and hundreds of thousands of galaxies with high redshifts, fainter than $ I_{\mathrm{AB}} = 22.0$ mag and without strong emission lines, this is practically impossible. Spectroscopy of such faint galaxies requires the largest telescopes and exposure times of several hours \citep {LeFevre2005, Gerke2005, Meneux2006, Cooper2006, Coil2007, Lilly2007}. 

There are several medium-band surveys of sufficient depth that allow us to solve statistical problems of studying the physical properties of galaxies: COMBO-17 \citep [Classifying Objects by Medium-Band Observations, a spectrophotometric 17-filter survey,] [] {Wolf2004}, ALHAMBRA \cite [Advanced Large, Homogeneous Area Medium Band Redshift Astronomical Survey,] [] {Moles2008}, COSMOS \citep [Cosmic Evolution Survey,] [] {Murayama2007}, miniJPASS (a set of J-PAS-like data for studies the scientific capabilities of J-PAS \citep[The Javalambre--Physics of the Accelerating Universe Astrophysical Survey][]{2009ApJ...691..241B, 2014arXiv1403.5237B}, \citealt*{2021arXiv210213121G}). Some of these surveys were performed on small-sized areas that are significantly spaced apart, which makes it difficult to study the physical properties of galaxies depending on the density of the environment (e.g. COMBO-17, ALHAMBRA). Broadband surveys, because of the low accuracy of determining photometric redshifts and spectral type classification of galaxies, are excluded from consideration. Spectral surveys of sufficient area, such as SDSS \citep[Sloan Digital Sky Survey,][]{Peng2010} and 2dFGRS \citep[The 2dF Galaxy Redshift Survey,][]{2003astro.ph..6581C}, are limited in-depth, while deep spectral surveys are insufficient in the area and samples are not complete due to the need for preliminary selection of objects. These are the reasons photometric surveys using medium-band filters are becoming increasingly relevant.

In 2013 -- 2015 the Laboratory of spectroscopy and photometry of extragalactic objects of
the Special Astrophysical Observatory together with Armenian specialists upgraded the 1-m Schmidt telescope of the Byurakan Astrophysical Observatory of the National Academy of Sciences of Armenia \citep{Dodonov2017}. We completely redesigned the control system of the telescope: we replaced the actuating mechanisms, developed telescope control software, and made the guiding system. We redesigned and prepared a 4k x 4k Apogee (USA) liquid-cooled CCD with RON $\sim$ 11.1 $\mathrm{e^{-}}$, a pixel size of 0.868, and a ﬁeld of view of about 1 $\mathrm{deg^2}$, and in October 2015 mounted it in the telescope's focus. The detector is equipped with a turret bearing 20 medium-band ﬁlters (FWHM = 250 \AA) uniformly covering the 4000 –- 9000 ~\AA\!\!~\AA $\;$ wavelength range, ﬁve broadband ﬁlters ($u$, $g$, $r$, $i$, $z$ SDSS), and three narrowband ﬁlters (5000 \AA , 6560 \AA $\;$ and 6760 \AA, FWHM = 100 \AA). The main programs of the telescope are the search for young stellar objects, the evolution of AGN, the stellar composition of galaxy disks, and the study of the evolution of the main characteristics of galaxies.

The progress in the modern physics of galaxies, associated with the growth of the number of observational data and the development of numerical modeling methods, has not yet completely clarified the issues of the formation and evolution of galaxies. They remain open and relevant. The variety of shapes and types of galaxies shows they evolve under the influence of a significant number of conditions: the density of the environment, the rate of accretion of external matter, internal secular evolution, the active nuclei feedback, etc. Each of these conditions affects the rate of star formation in the galaxy, which leaves an "imprint" on the history of star formation (SFH).

This work presents two parts of the gMOSS (Galaxies of Medium-band One-meter Schmidt Survey) photometric catalogue of 19,875 galaxies in a fixed aperture and 19,029 in Kron-like apertures, high-precision photometric redshifts from 1-m Schmidt telescope of Byurakan Astrophysical Observatory and spectral observation on the 6-meter BTA telescope of SAO RAS. Section \ref{s:data} briefly describes the data and observations. Photometry, redshift estimation, and observational properties of the sample are given in Section \ref{s:danalysis}. Section \ref{s:catalogue} comprises cluster galaxies catalogues descriptions. Estimating the physical parameters of galaxies and brief analysis of galaxy stellar populations are described in Section \ref{s:phys.prop.}. And we discuss our results in Section \ref{s:Discussion}

The paper uses the cold dark matter cosmological $\Lambda$CDM model with parameters $ \Omega_M = 0.315 $, $ \Omega_ \Lambda = 0.685 $ and $ \mathrm {H_0 = 67.4 \; km \; s ^ {- 1}} $ \citep{2018A&A...617A..48P}. All the stellar masses in this work are quoted in solar mass units ($\mathrm{M_{\odot}}$) and are scaled according to a universal \cite{2003PASP..115..763C} initial stellar mass function. All the magnitudes are in the AB system \citep{1983ApJ...266..713O}.

\section{Data and Observations}
\label{s:data}

\begin{figure}
\center{\includegraphics[width=1.0\linewidth]{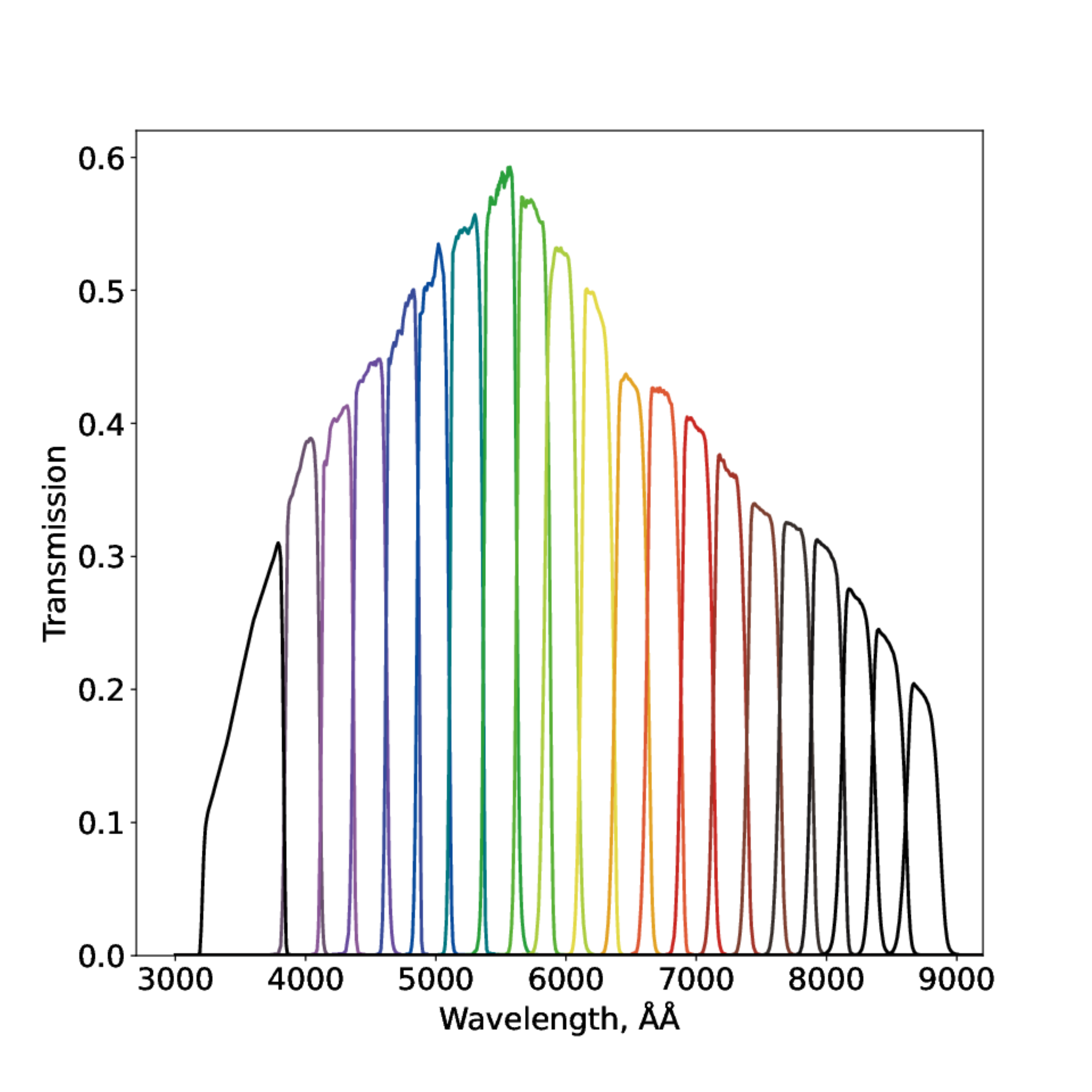}}
\caption{Set of 17 filters used for SED (Spectral Energy Distribution) construction. The leftmost passband corresponds to $u$ SDSS filter and the others are medium-band filters. Filter transmission was measured in F / 2, the spectral sensitivity of the CCD detector was taken into account.}
\label{ris:image1}
\end{figure}

In this work, we use the observational data of HS47.5-22 field \citep[Hamburg Quasar Survey,][]{Molthagen1997} obtained by the 1-meter Schmidt telescope \citep{Dodonov2017} of the NAS RA Byurakan Astrophysical Observatory during several sets in February, March, April, and November 2017 and February and November 2018 years.  The telescope field of view with 4k $\times$ 4k CCD is $58 \times 58$ arcmin, with a scale of $0.868 \; \mathrm{arcsec \; {pixel}^{-1}}$. Observations were made in 4 broadband filters ($u, g, r, i$ SDSS) and 16 medium-band filters (FWHM = 250 ~\AA ~\; with homogeneous coverage of the spectral range 4000 -- 8000 ~\AA\!\!~\AA). Medium-band with $u$ SDSS filters are presented in Fig. \ref{ris:image1}.

Four sets of exposures covered the central part of the field in broadband and medium-band filters. The overlap of adjacent sets was about 10 arcmin. The total exposure time was selected to reach a depth of $ m_{\mathrm{AB}} \approx 25$ mag with a signal-to-noise ratio of $ \sim 5 $ in broadband ($\sim$ 2 h) and $ m_{\mathrm{AB}} \approx 23$ mag with a signal-to-noise ratio of $ \sim 5 $ in medium-band filters (about 60 min at the peak of the detector sensitivity curve and about 2 h at the edges of the range). From these observations, we create a mosaic of  4 x 1 $\mathrm{deg^2}$ fields with a total area of 2.386 $\mathrm{deg^2}$. 

In addition, we carried out long-slit spectral observations of certain galaxies with the Russian 6-m telescope with SCORPIO-2 multi-mode reducer of the telescope prime focus \citep{Afanasiev2011}. The observations took place on 25, 28 of February 2020 (under the seeing 2.1 and 3.0 arcsec respectively) and 13, 14 of December 2020 (under the seeing 1.5 and 1.3 arcsec respectively). The slit width was 2 arcsec. We used the diffraction grating VPHG940@600, which covers the spectral range 3500 -- 8500 ~\AA\!\!~\AA $\;$ and has a dispersion of 1.16 ~\AA $\;$ pixel$^{-1}$. The spectral resolution is $\sim$ 7.0 ~\AA $\;$ (R $\sim$ 5200) estimated as FWHM for 1 arcsec slit of night-sky emission lines. The total exposure time was selected to get a signal-to-noise ratio of $ \sim$ 5 -- 10.

We processed the spectra of all objects that hit the slit and had a sufficient signal-to-noise ratio. We got the spectra of 29 galaxies. We used a standard processing software package for spectra obtained by multi-mode focal reducer SCORPIO-2. The data reduction steps include bias subtraction, cosmic particle removing, flat-field correction, wavelength calibration, sky subtraction, correction for the atmospheric and spectrograph transparency by spectrophotometric standards, and extraction to 1D spectrum.  The accuracy of the spectral redshift estimation for all galaxies is 0.002.

All data is available to the reviewer(s) and will be made open-source on publication.

Also, we use spectral and photometric data from the SDSS database (Sloan Digital Sky Surveys, Data Release 16, \cite{2020ApJS..249....3A}) for calibration of photometric redshifts and GAIA DR2 \cite{2016A&A...595A...1G, 2018A&A...616A...1G} to exclude objects with proper motions (like stars). 

\section{Observation data analysis}
\label{s:danalysis}
This section presents a summary of the observation data analysis methods used in the gMOSS survey.

\subsection{Photometry}

Photometry of the objects was obtained using \textsc{sextractor} \citep{Bertin1996} in dual image mode. The base image was created from the sum of deep ($\sim 25$ mag) images obtained in $g, r$ and $i$ SDSS filters. Before the photometry, all images were convolved to common seeing quality and transformed to a common coordinate system. For galaxy type classification and redshift estimation, we use photometry in fixed apertures MAG\_APER and correct received fluxes for light loss using a light curve obtained from bright stars. The diameter of the aperture was 3 arcsec. Using a fixed aperture allows us to reduce the number of objects with strongly biased aperture photometry by neighboring sources. For galaxy physical parameter estimation we received Kron-like fluxes for 97 per cent galaxies of the total sample in the field by using MAG\_AUTO \textsc{sextractor} photometry because of large photometric errors for some faint sources. Photometric calibration was developed using spectral and photometric data from the SDSS survey for the objects detected in the field. By using field objects as standard stars within each exposure, we were independent of photometric conditions during imaging.


The galaxy sample of the HS47.5-22 field is limited by the threshold magnitude $ m_{\mathrm{AB}} = 22.5$ mag in the $r$ SDSS filter with redshift $z \le 0.8$. The total number of field objects down to $ m_ {\mathrm{AB}}$ $\sim 25$ mag is approximately 85,000; the sample of galaxies intended for the study included 18,079 galaxies with photometric data in a fixed aperture and 17,545 galaxies with the use of Kron-like apertures applicable to the selection criteria (see Section \ref{ss:sample}).

\begin{figure}
\includegraphics[width=1.1\linewidth]{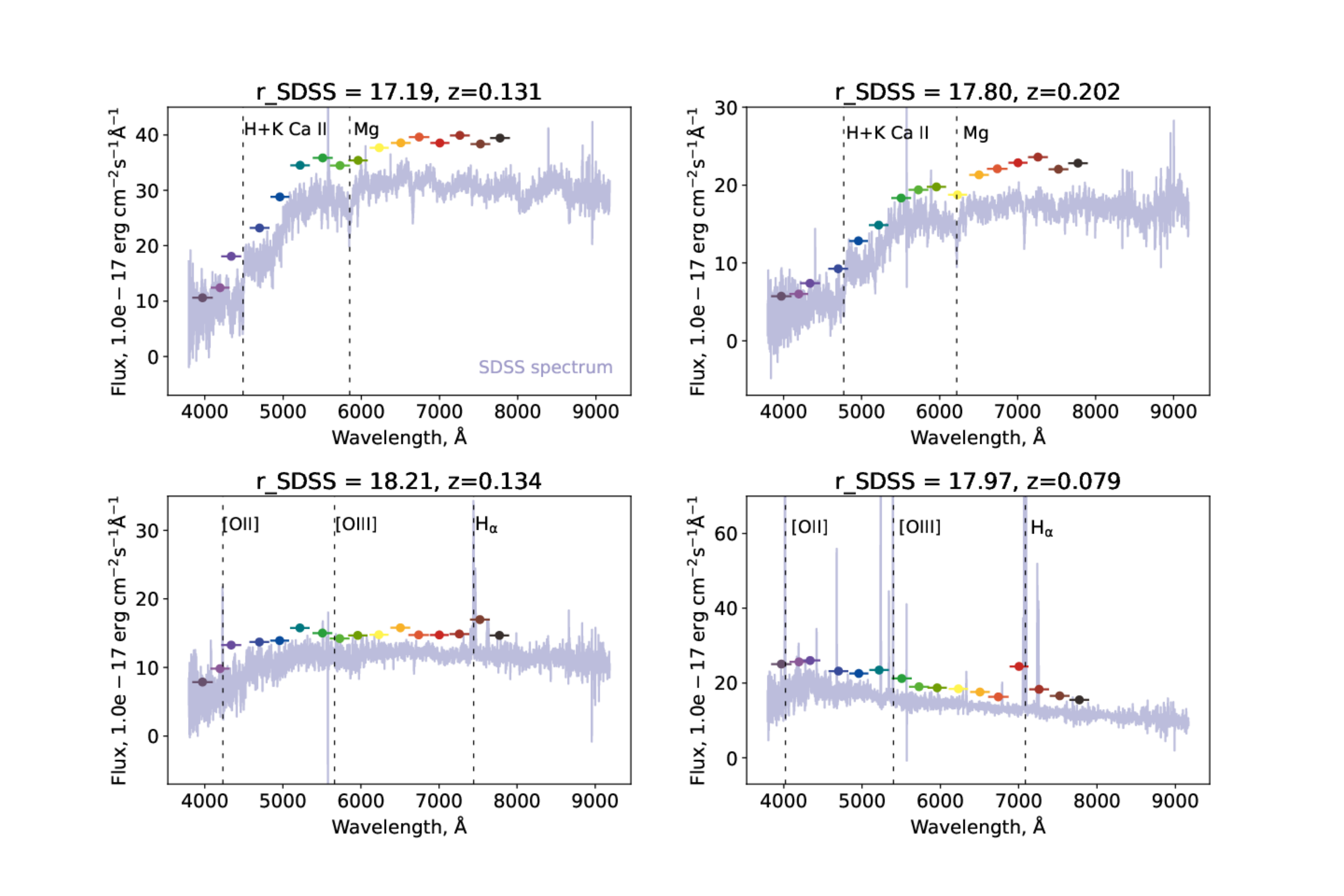}
\caption{SED for 4 galaxies from HS47.5-22 ﬁeld. The horizontal bar is the ﬁlter bandwidth, and the solid line is the spectra of these objects from the SDSS survey. Flux differences seen in the picture are because of the limited size of the fibers (3 arcsec) used for spectroscopy. The main spectral features (e.g. H+K Ca \textsc{ii}, Mg, $\mathrm{H \; \alpha}$, [O \textsc{ii}] and [O \textsc{iii}] - dashed lines) of each object are also clearly visible in the SEDs.}
\label{ris:image2}
\end{figure}

Photometric measurements from 17 filters provide low-resolution spectra for each object which are analyzed by a statistical technique for classification and redshift estimation based on spectral template matching. SEDs for 4 galaxies from HS47.5 - 22 ﬁeld in 16 medium-band filters and their SDSS spectra are shown in Fig.\ref{ris:image2}. The main spectral features such as H+K Ca \textsc{ii}, Mg, $\mathrm{H \;\alpha}$, [O \textsc{ii}], [O \textsc{iii}] and others are clearly visible in the SEDs. The photometric properties of a galaxy sample of the HS 47.5-22 field were studied in the range from 4000 ~\AA \; to 8000 ~\AA, which makes it possible to determine the redshifts of galaxies up to $z = 0.8$.

\subsection{Photometric redshifts}

The method for determining the photometric redshift and SED type of galaxies is based on the correspondence of the spectral templates of the galaxy to the observed energy distribution. We used the galaxy's spectral templates library from \cite{Dodonov2008}, and a set of programs \textsc{zebra} \citep[Zurich's Extragalactic Bayesian Redshift Analyzer,][] {Feldmann2006}. 

\begin{figure}
\center{\includegraphics[width=1.0\linewidth]{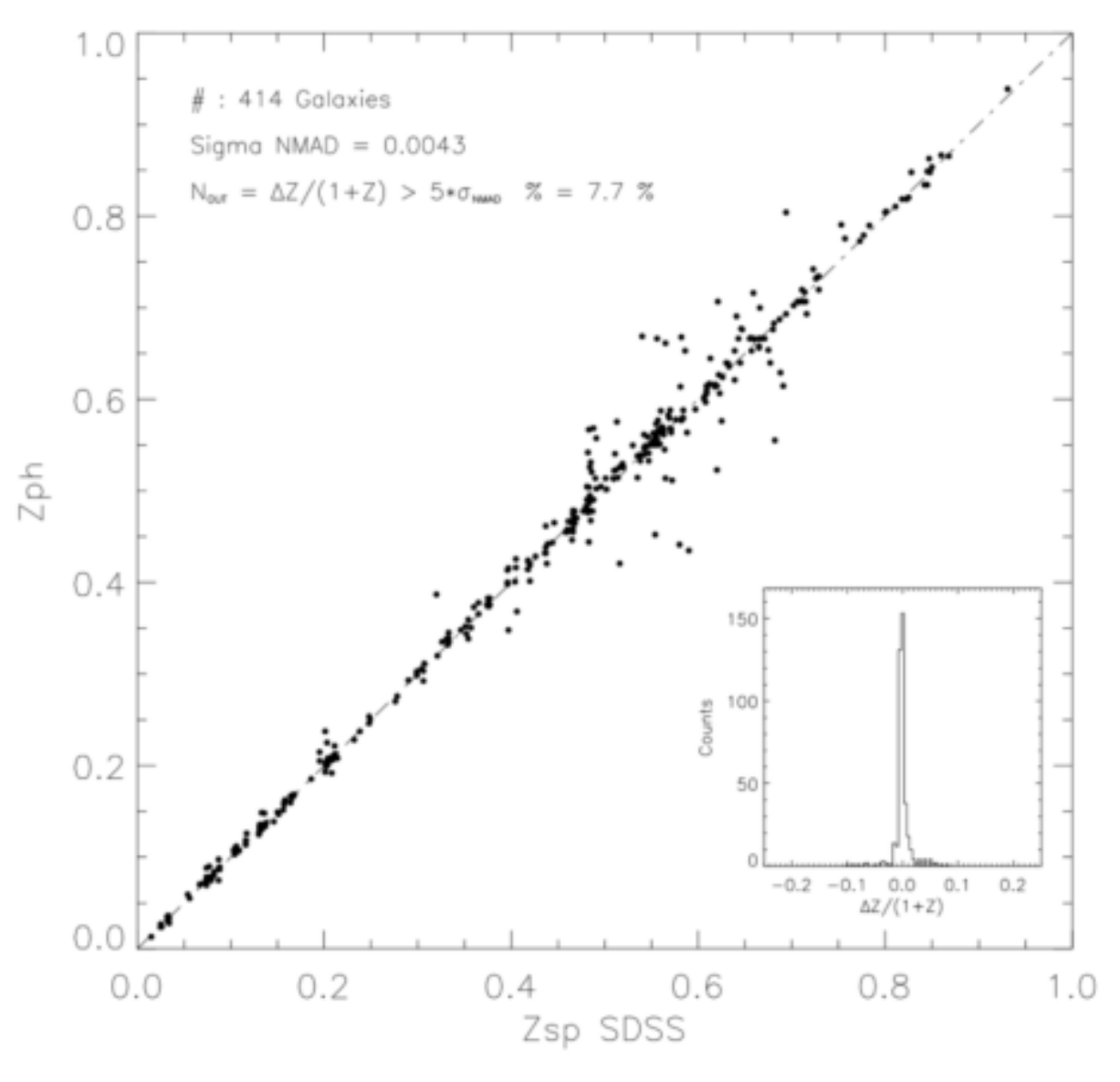}}
\caption{Comparison of the photometric redshifts $ z_ \mathrm {ph} $ of the galaxies obtained with the \textsc{zebra} software tool with the spectroscopic redshifts $ z_ \mathrm {sp} $ of the galaxies taken from SDSS for 414 galaxies with spectroscopy. The accuracy of determining the photometric redshift is $ \sigma_\mathrm{{NMAD}} <0.0043 $, the outlier's percentage is $ \Delta z / (1 + z)> 5. * \sigma_\mathrm{{NMAD}} \sim 7.7$ per cent. Accuracy $\sigma_\mathrm{{NMAD}}$ changes from 0.002 for objects brighter than 19 mag in $r$ SDSS filter and 0.006 for objects brighter than 21 mag till 0.007 for objects brighter than 22.5 mag. The dashed line denotes a line of equal photometric and spectral redshifts.}
\label{ris:image3}
\end{figure}

The obtained redshift accuracy is $\sigma_\mathrm{{NMAD}} < 0.0043$ and the fraction of catastrophic outliers is ($\Delta z/(1+z) > 5.\cdot\sigma_\mathrm{{NMAD}}$) $\sim 7.7$ per cent. Figure \ref{ris:image3} shows a comparison of the photometric redshifts of galaxies obtained using the \textsc{zebra} software tool and the spectroscopic redshifts from the SDSS database. The total number of galaxies with spectroscopic redshifts in the observed HS47.5-22 field is 414. Accuracy $\sigma_\mathrm{{NMAD}}$ changes from 0.002 for objects brighter than 19.0 mag in $r$ SDSS filter and 0.006 for objects brighter than 21.0 mag till 0.007 for objects brighter than 22.5 mag.

\subsection{Galaxy sample}
\label{ss:sample}

A sample of galaxies for the study was made from a complete photometric survey (approximately 85,000 objects down to $ m_{\mathrm{AB}}$ $\sim 25$ mag) according to the following criteria:
\begin {enumerate}
\item To exclude objects with proper motions, such as stars, we used data from the GAIA DR2 survey \citep{2016A&A...595A...1G, 2018A&A...616A...1G};
\item To exclude stars (without proper motions in the GAIA survey), we remove objects for which the morphological classification from the DECALS survey \citep{2019AJ....157..168D} is STAR;  
\item To exclude objects with strongly biased aperture photometry by neighboring sources or by many bad pixels in any aperture we include to the sample only objects with the automatic index of contamination of \textsc{sextractor}, as provided by the FLAGS$\leq 2$;
\item To define a flux-limited sample, we imposed the condition that the galaxies have to be brighter than  $r_{\mathrm{AB}} = 22.5$ mag; 

\end {enumerate} 

Applying this set of criteria to the initial sample  of over 85,000 objects, we got a sample of galaxies, selected according to $r$ SDSS magnitude, extended and contamination index. Our final sample of objects with aperture photometry contains 19,875 extended and non-contaminated sources with $r_{\mathrm{AB}} \le 22.5$. For Kron-like photometry, we obtain 19,029 objects because of large photometric errors for some faint sources. MAG\_AUTO provides the total flux of the source within an elliptical aperture determined by the KRON\_RADIUS, MAG\_APER measure for a smaller fixed aperture (3 arcsec).

It should be noted that the final gMOSS survey contains galaxies with a wide variety of spectrum types, colours, and environments (see Fig. \ref{ris:image2} and Grokhovskaya et.al., 2022, in prep.). The SEDs of galaxies on upper panels of Fig. \ref{ris:image2} shows the power of gMOSS to identify red galaxies, to detect the absorption lines such as H+K Ca \textsc{ii} and Mg. Blue and star-forming galaxies with the recombination nebular lines $\mathrm{H \; \alpha}$ and the other nebular collisional lines, such as [O \textsc{ii}] $\lambda3727$ and [O \textsc{iii}] $\lambda5007$ are also very well-identified in the survey (see the bottom panel in Fig. \ref{ris:image2}). Galaxies in groups, as well as galaxies in less dense environments, are well identified because we use a fixed 3 arcsec aperture to identify objects. It allows us to analyze the dependence of the physical properties of galaxies on the density of environments in the next work.

\begin{figure*}
\begin{minipage}[H]{0.33\linewidth}
\center{\includegraphics[width=1.0\linewidth]{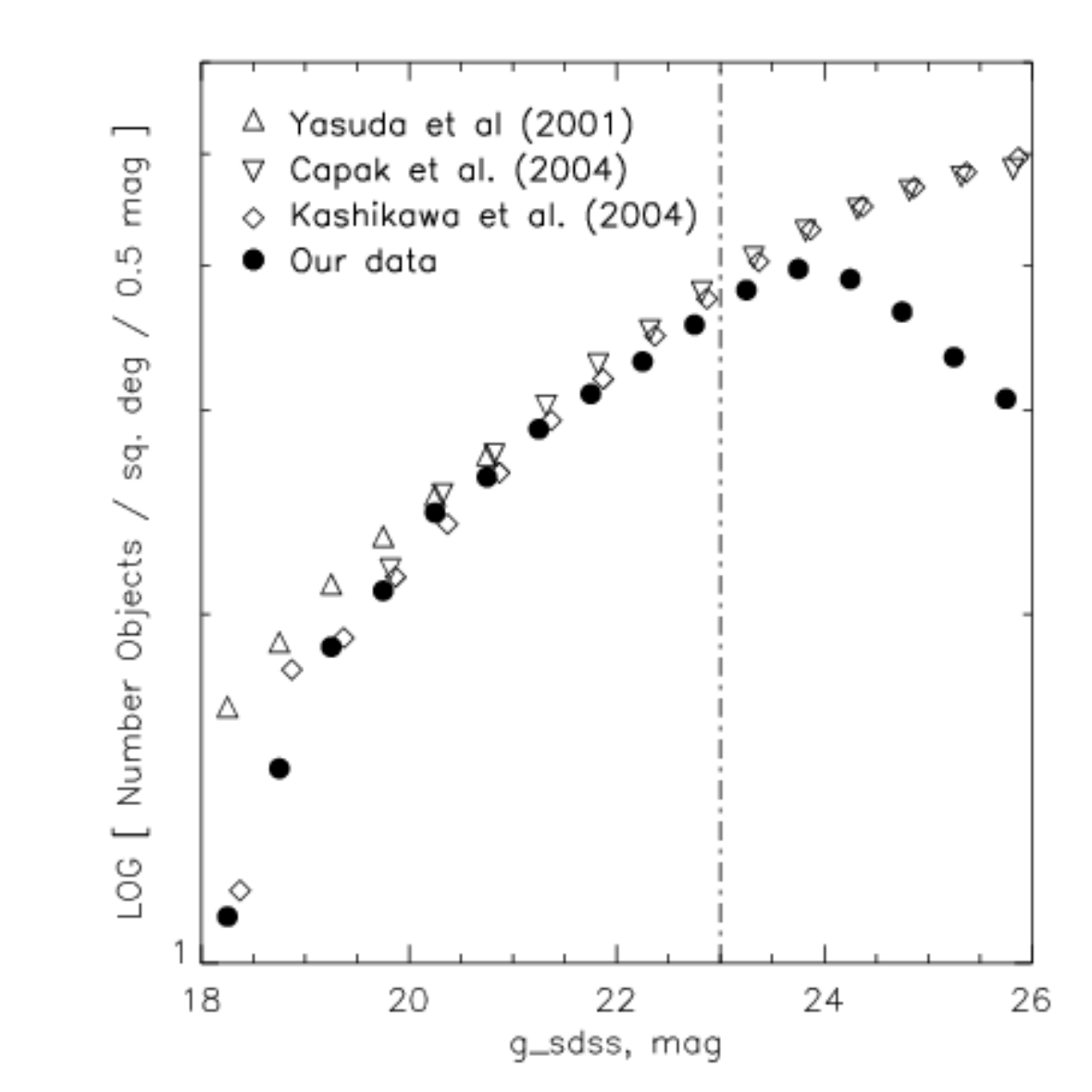}}
\end{minipage}
\hfill
\begin{minipage}[H]{0.33\linewidth}
\center{\includegraphics[width=1.0\linewidth]{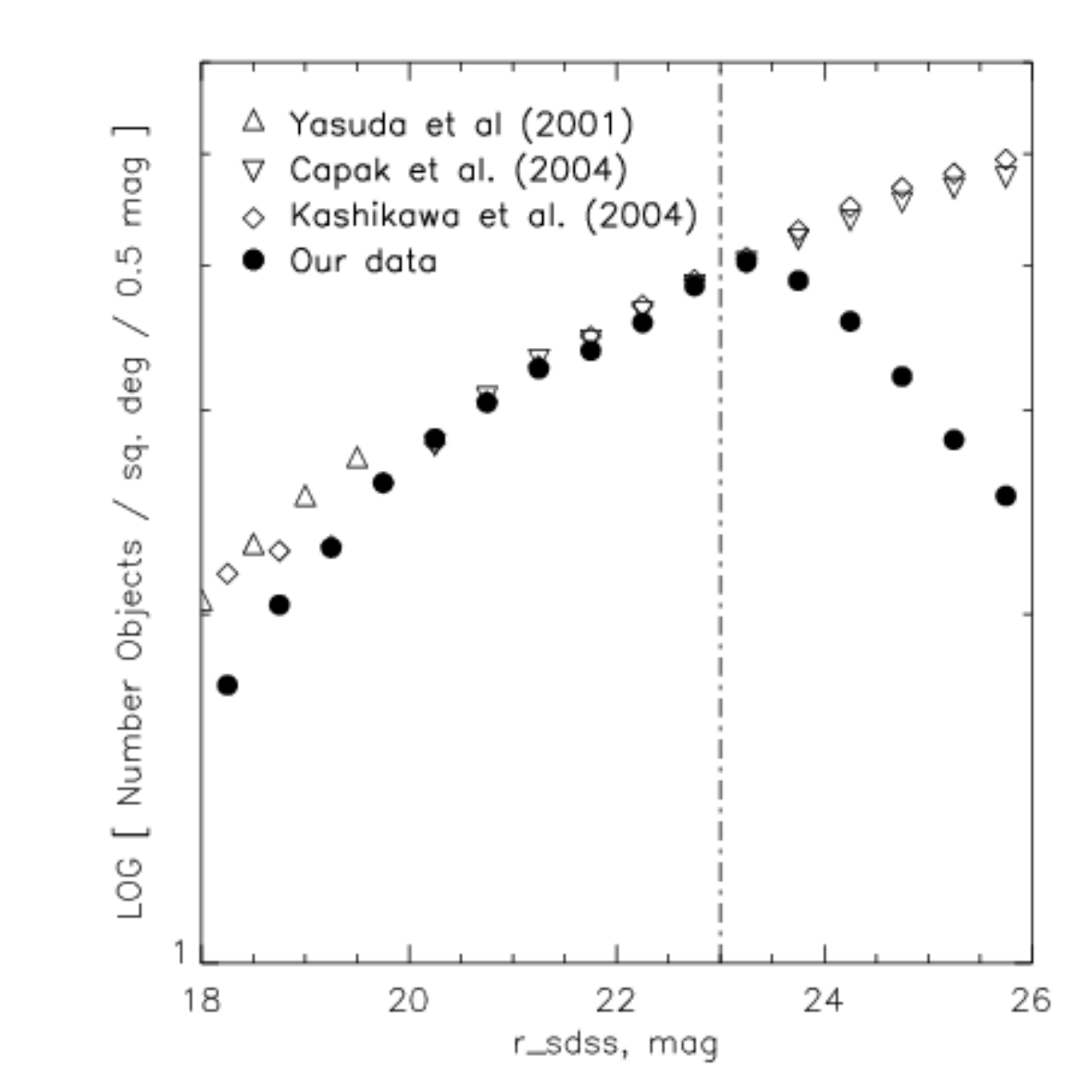}}
\end{minipage}
\hfill
\begin{minipage}[H]{0.33\linewidth}
\center{\includegraphics[width=1.0\linewidth]{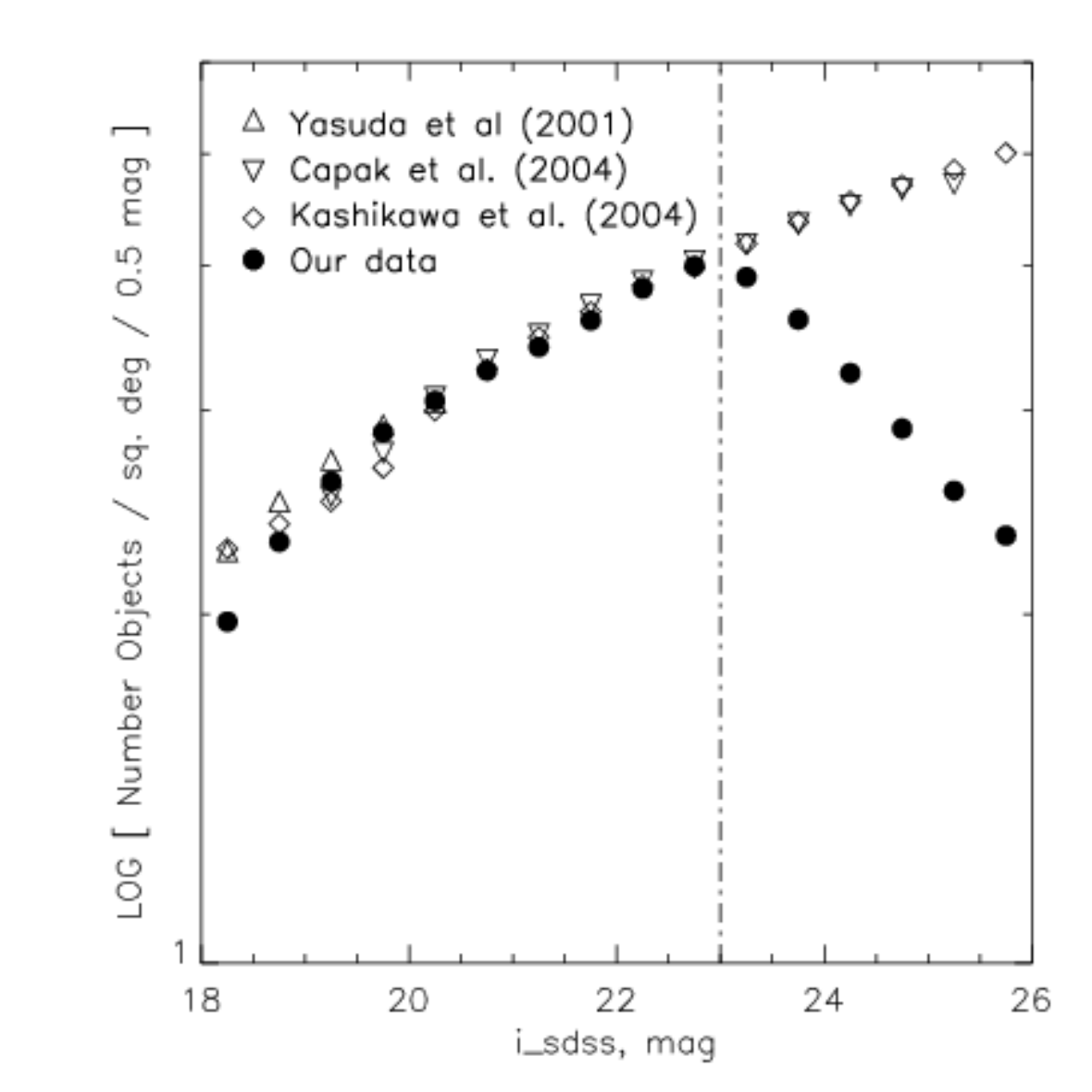}}
\end{minipage}
\caption{Comparison of the completeness of the galaxy sampling in the $g$, $r$, and $i$ SDSS filters in the previously published data and the data obtained in this work. The figure shows the lack of colour selection effects down to $m_{\mathrm{AB}}=23$ mag.}
\label{ris:image4}
\end{figure*}

The completeness of the sample of galaxies number-counts in $g$, $r$ and $i \; SDSS$ filters was verified by comparison with previously published data in the papers \cite{2001AJ....122.1104Y, 2004AJ....127..180C, 2004PASJ...56.1011K}. Results are shown in Fig. \ref{ris:image4}. The galaxy sample is complete up to $m_\mathrm{AB}=23.0$ mag with no colour selection effects in the all-optical range. 

\subsection{Observational properties of the sample}

\begin{figure}
\includegraphics[width=1.1\linewidth]{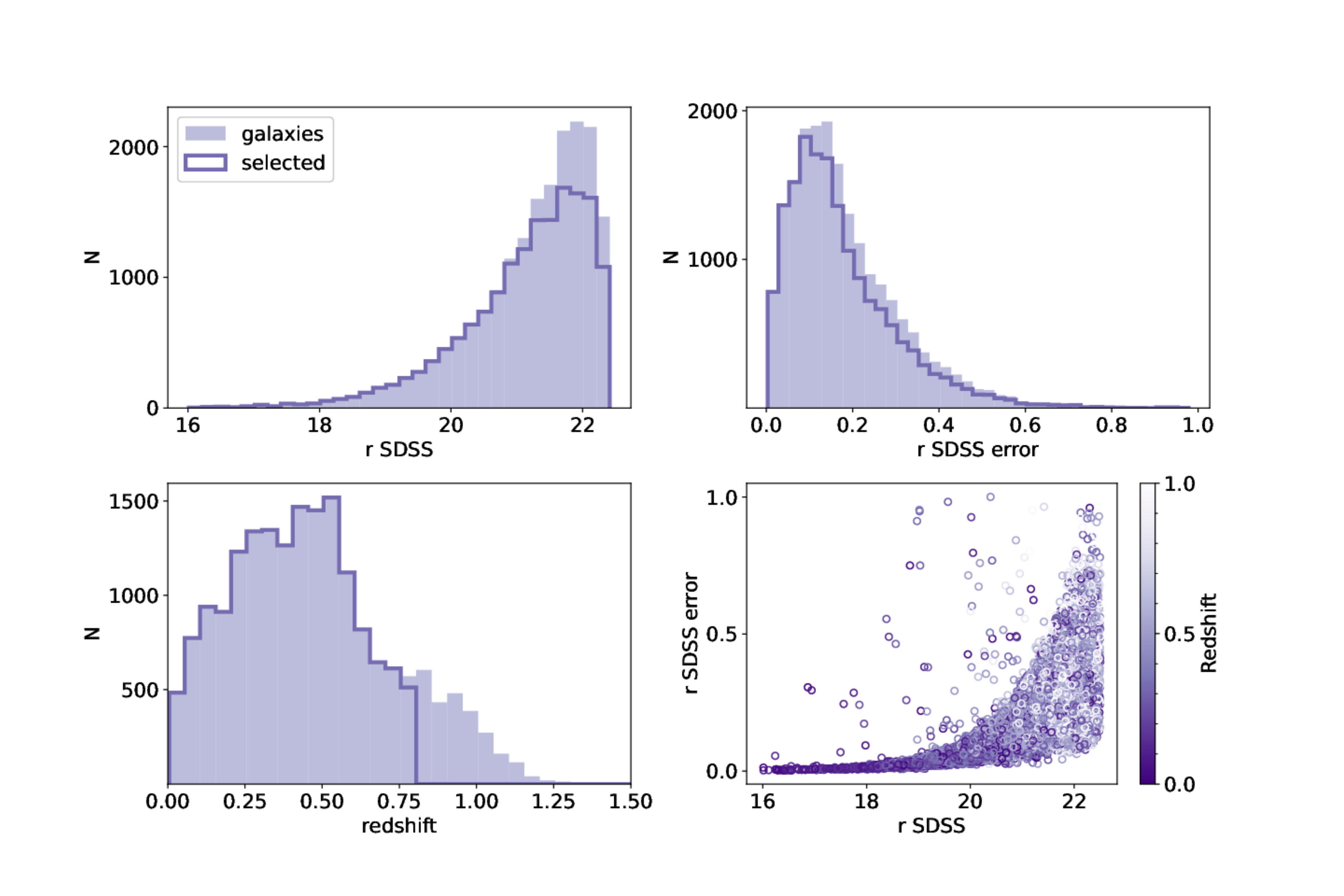}
\caption{Distributions of several observational properties. \emph{Top-left, top-right and bottom-left panels}: Distribution of $r$ SDSS magnitude and errors, and redshifts of galaxies with $r$ SDSS $\le 22.5$ mag identified in gMOSS (light purple) and galaxies with $z\le 0.8$ (dark purple) and photometric redshifts $z \le 0.8$. All magnitudes and related errors were obtained using \textsc{sextractor}. \emph{Bottom-right}: Errors of the $r$ SDSS as a function of $r$ SDSS band magnitude. The colour bar illustrates the redshift of each galaxy.}
\label{ris:image5}
\end{figure}

Figure \ref{ris:image5} compares the distributions of redshift, magnitude, and error in the $r$ SDSS band of the full sample of gMOSS galaxies and the final selected sample of galaxies up to $ z = 0.8 $. As seen from the top-right panel of Fig. \ref{ris:image5}, the redshift sampling limit excludes most of the galaxies with significant errors in determining the magnitude in the $r$ SDSS filter, as well as some of the faint objects in the sample. After the redshift $z=0.3$, we observe a sharp increase in the number of objects associated with a rapid increase in the observed volume. After the redshift $ z = 0.6 $, there is a decrease in the number of objects because of the selected threshold on the magnitude of $r_{\mathrm{AB}} \le 22.5$. Most of the galaxies in the sample have a photometric redshift between 0.2 and 0.6. 

\begin{figure}
\includegraphics[width=1.1\linewidth]{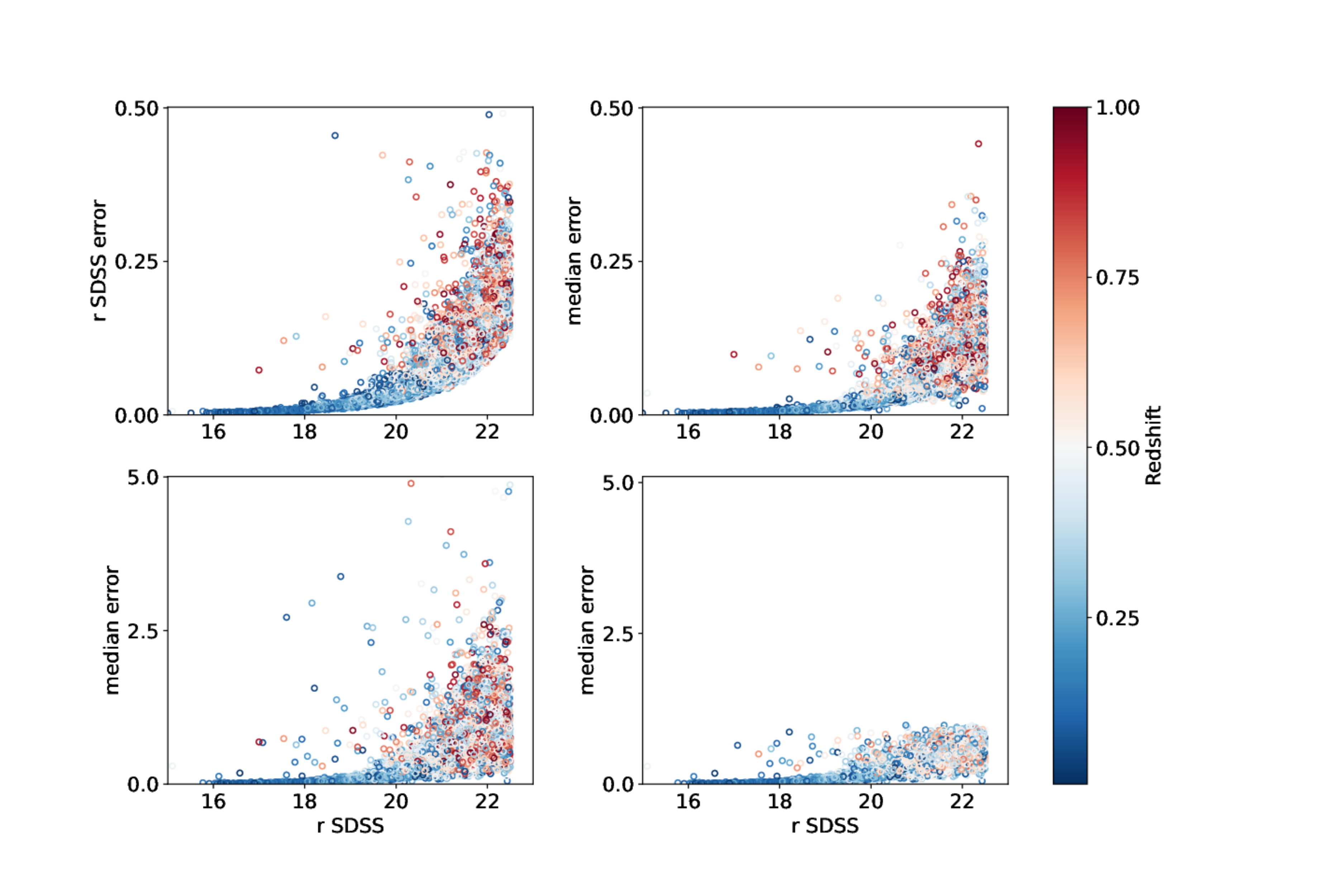}
\caption{Errors of the $r$ SDSS band (\emph{Top left panel}) and medium-band filters (\emph{Top right panel}) as a function of the $r$ SDSS magnitude. The \emph{top panels} show the aperture photometry. The \emph{bottom panels} show the Kron-like photometry and errors of the medium-band photometry. \emph{Bottom left panel}: full sample of galaxies, \emph{Bottom right panel}: part of the complete sample, which is used to analyze the properties of the stellar population of galaxies. The colour bar illustrates the redshift of each galaxy.}
\label{ris:image6}
\end{figure}

Figure \ref{ris:image6} shows the correlation between the magnitude of objects in the $r$ SDSS filter and photometry errors, both in the $r$ SDSS filter and for the average error in the medium-band filters. For photometry in Kron-like apertures, we obtained a strong increase in the average error in the medium-band filters for faint objects. This is because faint objects ($r$ SDSS $\sim 20.5$ mag) in the part of the medium-band filters have an insufficient signal-to-noise ratio (S/N $\le 5$) for the correct operation of the photometry algorithm with Kron-like apertures. The \emph{bottom left panel} shows galaxies with a Kron-like aperture, which are used to get the properties of stellar populations of galaxies using the SED-fitting code \textsc{cigale}  \citep{ 2005MNRAS.360.1413B, 2009A&A...507.1793N, 2019A&A...622A.103B}. The conditions for using the data in the Kron-like apertures were that the median error of the photometry was less than 1 mag and at least half of the filters were with a sufficient signal-to-noise ratio (S/N $\ge 5$) for detecting an object by the Kron-like aperture detection algorithm. Thus, we could use Kron-like photometry data for 9620 galaxies from a complete sample. For the rest of the galaxies, we used aperture photometry data. This is true for estimating distant faint objects whose dimensions do not exceed 3 arcsec.

\begin{figure}
\includegraphics[width=1.1\linewidth]{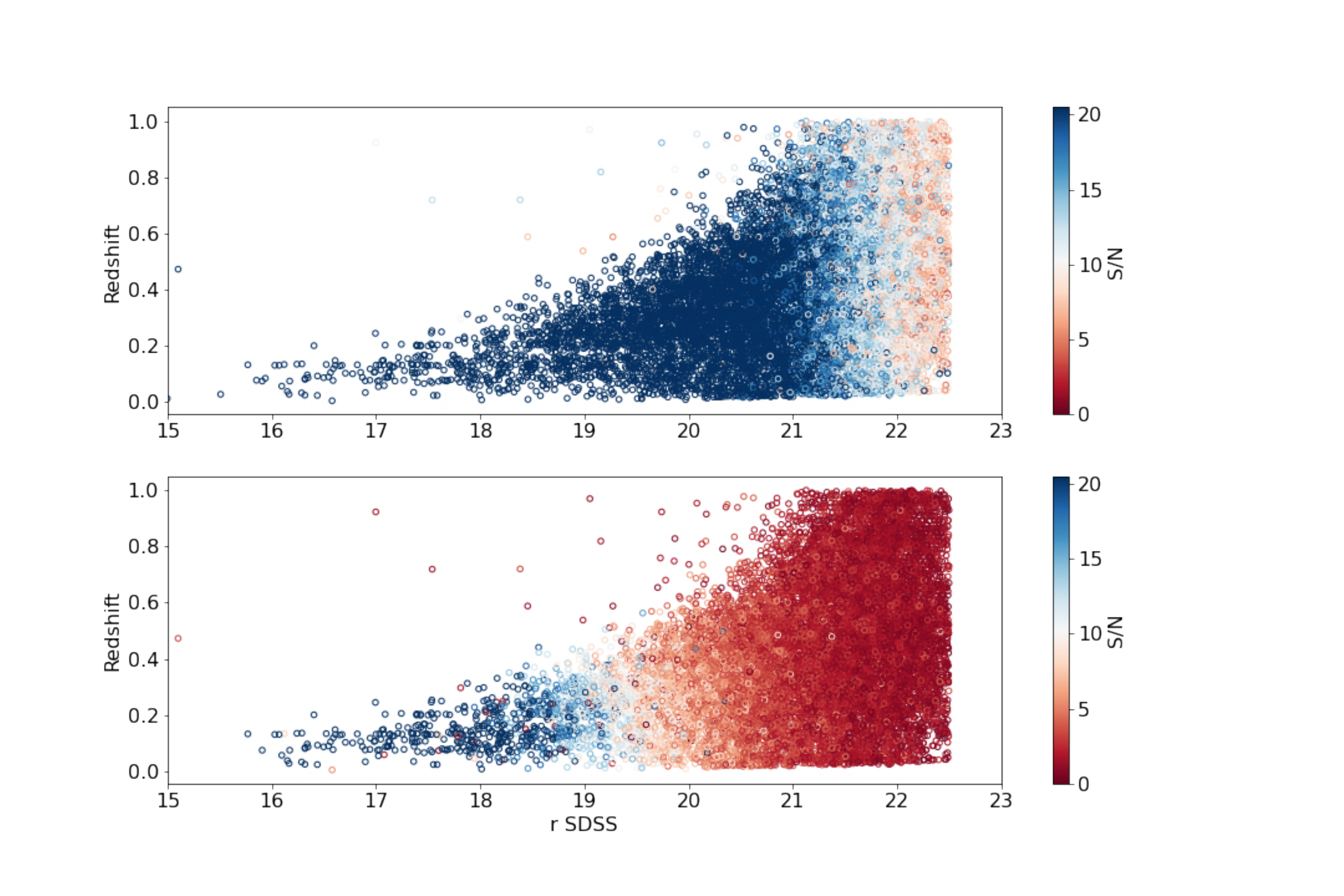}
\caption{Galaxy observed magnitudes in the $r$ SDSS filter as a function of redshift for fixed aperture and Kron-like photometry (\emph{Top} and \emph{bottom panels}, respectively). The colour bar shows the median S/N ratio in the medium-band filters.}
\label{ris:image7}
\end{figure}

The magnitude in the $r$ SDSS band shows a dependence on redshift Fig. \ref{ris:image7}. Galaxies with $r$ SDSS $\le 20.0$ mag (fixed aperture photometry) are often at $z < 0.5$, while fainter galaxies are at any distance. The S/N ratio is also a clear function of the brightness of the galaxy, as indicated by the magnitude in the $r$ SDSS filter and median S/N ratio in the medium-band filters. The increase in the S/N for Kron-like aperture photometry regarding fixed aperture photometry is clear in Fig. \ref{ris:image7}, although this increase is quite significant. For instance, galaxies with $r$ SDSS $\sim 21.0$ mag have an S/N $\sim 10$ for fixed aperture apertures, while for Kron-like apertures this value is $\sim 3$. Therefore, for galaxies with an insufficient signal-to-noise ratio for Kron-like photometry (S/N $\le 5$, typically these are galaxies fainter than 20.5 mag in $r$ SDSS filter), we use fixed-aperture photometry data for analysis of galaxy physical properties.

We can fit $\sim 74$ per cent of the full gMOSS galaxy survey (in a combination of aperture photometry and Kron-like photometry data) with the \textsc{cigale}  \citep{ 2005MNRAS.360.1413B, 2009A&A...507.1793N, 2019A&A...622A.103B} SED-fitting code  with an acceptable parameter value of reduced $\chi^{2}$ (see Section \ref{s:phys.prop.}). The unfitted spectra do not fulfill quality requirements imposed by the SED-fitting codes, such as the minimum S/N in each band. We select the total exposure time to get a signal-to-noise ratio of $\sim 5-10$  for objects brighter than $r$ SDSS = 22.5, however, for some faint objects this time may not achieve the required S/N ratio.

\section{Catalogue description}
\label{s:catalogue}

We prepare two catalogues that were used for the analysis of stellar populations and photometric redshifts of galaxies: photometry of galaxies in fixed apertures (aperture size 3 arcsec) and Kron-like apertures as two parts of gMOSS galaxy catalogue. Both catalogues have the same structure as shown in Table \ref{tab:cat_description}. The columns of the table contain the following information about the catalogue: (1) - parameter name, (2) - unit of the parameter (3) - description of the parameter. The magnitudes are in the AB magnitude system and are not corrected for Galactic extinction.

\begin{table}
	\centering
	\caption{Column Description of Catalogues}
	\label{tab:cat_description}
	\begin{tabular}{lll} 
  \hline
  Column                & Unit              & Description                    \\
  \hline
  id$\_$gal             & -                 & Unique object ID                  \\
  ra                    & degree            & R.A. in J2000                   \\
  dec                   & degree            & decl. in J2000                   \\
  zph             & -                 & Estimated photometric redshift           \\
  zph\_{err}      & -                 & Estimated photometric redshift error        \\
  zsp               & -                 & Estimated spectral redshift            \\
  m400                  & mag               & mb400-band magnitude                \\
  er400                 & mag               & mb400-band magnitude error             \\
  m425                  & mag               & mb425-band magnitude                \\
  er425                 & mag               & mb425-band magnitude error             \\
  m450                  & mag               & mb450-band magnitude                \\
  er450                 & mag               & mb450-band magnitude error             \\
  m475                  & mag               & mb475-band magnitude                \\
  er475                 & mag               & mb475-band magnitude error             \\
  m500                  & mag               & mb500-band magnitude                \\
  er500                 & mag               & mb500-band magnitude error             \\
  m525                  & mag               & mb525-band magnitude                \\
  er525                 & mag               & mb525-band magnitude error             \\
  m550                  & mag               & mb550-band magnitude                \\
  er550                 & mag               & mb550-band magnitude error             \\
  m575                  & mag               & mb575-band magnitude                \\
  er575                 & mag            & mb575-band magnitude error             \\
  m600                  & mag            & mb600-band magnitude                \\
  er600                 & mag            & mb600-band magnitude error             \\
  m625                  & mag            & mb625-band magnitude                \\
  er625                 & mag            & mb625-band magnitude error             \\
  m650                  & mag            & mb650-band magnitude                \\
  er650                 & mag            & mb650-band magnitude error             \\
  m675                  & mag            & mb675-band magnitude                \\
  er675                 & mag            & mb675-band magnitude error             \\
  m700                  & mag            & mb700-band magnitude                \\
  er700                 & mag            & mb700-band magnitude error             \\
  m725                  & mag            & mb725-band magnitude                \\
  er725                 & mag            & mb725-band magnitude error             \\
  m750                  & mag            & mb750-band magnitude                \\
  er750                 & mag            & mb750-band magnitude error             \\
  m775                  & mag            & mb775-band magnitude                \\
  er775                 & mag            & mb775-band magnitude error             \\
  mu                    & mag            & $u$ SDSS magnitude                  \\
  eru                   & mag            & $u$ SDSS magnitude error               \\
  mg                    & mag            & $g$ SDSS magnitude                  \\
  erg                   & mag            & $g$ SDSS magnitude error               \\
  mr                    & mag            & $r$ SDSS magnitude                  \\
  err                   & mag            & $r$ SDSS magnitude error               \\
  mi                    & mag            & $i$ SDSS magnitude                  \\
  eri                   & mag            & $i$ SDSS magnitude error               \\
		\hline
	\end{tabular}
\end{table}

Catalogues include unique object IDs, R.A. and Dec. coordinates (J2000), photometric measurements in 16 medium-band and 4 broadband SDSS filters, and their errors for each object and the accurate photometric redshifts ($\sigma_\mathrm{{NMAD}} <0.0043$ for all types of galaxies in full magnitude range) obtained from fixed aperture photometry data with spectral template matching by using a set of programs \textsc{zebra} \citep{Feldmann2006}. The unique object ID is a combination of the field number in the mosaic and the unique object number in each field (NB: the numbers of objects in different fields may overlap, so we additionally use the field identifier). In addition, we included data on spectroscopic redshifts from the SDSS survey for those galaxies for which this was possible \citep{2020ApJS..249....3A} and data on spectroscopic redshifts of 29 galaxies obtained with a 6-meter Russian telescope. The catalogues contain galaxies brighter than 22.5 mag in $r$ SDSS filter.

The total number of galaxies in the catalogue with photometric data in a fixed aperture is 18,079 galaxies, with the use of Kron-like apertures - 17,545. The number of objects in the catalogue with Kron-like apertures is slightly less than the number of objects in the catalogue with a fixed aperture  because this method does not separate close objects well.

In a separate table, we publish the physical properties of galaxies such as age of the main stellar population, star formation rate (SFR), luminosity, gas mass, stellar mass, and fitting quality parameter reduced $\chi^2$ obtained by using the SED-fitting code \textsc{cigale} \citep{ 2005MNRAS.360.1413B, 2009A&A...507.1793N, 2019A&A...622A.103B} with cross-ID of galaxies in photometry catalogues. The names of the parameters and units of measurement are presented in Table \ref{tab:tab_description}.

\begin{table*}
	\centering
	\caption{Column Description of Table with physical properties of galaxies}
	\label{tab:tab_description}
	\begin{tabular}{lll} 
  \hline
  Column & Unit & Description\\
  \hline
  id$\_$gal & - & Unique object ID\\
  chi$\_$sq & - & Reduced $\chi^2$ for galaxy SED\\
  param.restframe\_u\_prime-r\_prime& mag & Rest-frame colour $(u-r)_{res}$ \\
  param.restframe\_u\_prime-r\_prime\_err& mag & Rest-frame colour $(u-r)_{res}$ error \\
  stellar.age\_m\_star\_log & log [Myr] & Mass–weighted age \\
  stellar.age\_m\_star\_log\_err & log [Myr] & Mass–weighted age error \\
  attenuation.E\_BVs & mag & E(B-V)s, the colour excess of the stellar light for both\\
  &&the young and old population \\
  attenuation.E\_BVs\_err & mag & E(B-V)s error\\
  stellar.m\_star\_log & log [$M_{\odot}$] & Total stellar mass\\ 
  stellar.m\_star\_log\_err & log [$M_{\odot}$] & Total stellar mass error\\
  sfh.age\_main\_log   & log [Myr]              & Age of the main stellar population\\
                        &                   & in the galaxy\\
  sfh.age\_main\_log\_err  & log [Myr]               & Age of the main stellar population\\
                        &                   & in the galaxy error\\
  stellar.metallicity& dex & Metallicity\\
  sfh.sfr               & $\mathrm{M_{\odot}}$           & Instantaneous SFR\\
  sfh.sfr\_err           & $\mathrm{M_{\odot}}$           & Instantaneous SFR error\\
		\hline
	\end{tabular}
\end{table*}

In addition, in a separate table, we publish galaxies with spectral redshifts which were obtained with the Russian 6-meter telescope BTA and SCORPIO-2 multi-mode focus reducer. The columns of the table contain the following information about observations: (1) - unique galaxy id, (2), (3) - coordinates RA and DEC in degrees, (4) - spectroscopic redshift, (5) - exposition of observation. 

All data is available to the reviewer(s) and will be made open-source on publication.

\section{Stellar populations properties of galaxies in the sample}
\label{s:phys.prop.}
This section describes the distributions of the stellar population properties obtained with \textsc{cigale} SED-fitting code for the Kron-like MAG\_AUTO photometry of gMOSS catalogue.

To analyze the physical properties of stellar populations of galaxies, we used only objects up to $z \sim 0.8$, since most of the extended objects at $ z \ge 0.8 $ are fainter than 22.5 mag and are barely detected, and we do not have a sufficient number of objects for statistical analysis. The total number of galaxies for stellar populations properties estimation is 16,509.

\subsection{SED-fitting code parameters}

There are many modern codes for SED-fitting spectra in the full spectrum range from X-rays to radio such as \textsc{prospect} \citep{2020MNRAS.495..905R}, \textsc{beagle} \citep{2016MNRAS.462.1415C}, \textsc{bagpipes} \citep{2018MNRAS.480.4379C}, \textsc{cigale} \citep{ 2005MNRAS.360.1413B, 2009A&A...507.1793N, 2019A&A...622A.103B}, \textsc{prospector} \citep{Leja_2017, 2020zndo...4586953J}, \textsc{magphys} \citep{2008MNRAS.388.1595D}, \textsc{baysed} \citep{2012ApJ...749..123H, 2014ApJS..215....2H, 2019ApJS..240....3H} and others.  The stellar mass, rest-frame colours, ages, and other physical parameters in this work are derived using the \textsc{cigale} code. \textsc{cigale} (Code Investigating GALaxy Emission) is a multi-wavelength spectral energy distribution (SED) fitting code for extragalactic studies. The code has been developed to study the evolution of galaxies by comparing modeled galaxy SEDs to observed ones from the far X-rays to the radio.

Through template fitting, a series of stellar population parameters can be obtained, such as the rest-frame $(u - r)_{\mathrm{res}}$ colour, the mass-weighted age, V-band attenuation $A_\mathrm{v}$, total stellar mass, age of the main stellar population in the galaxy, metallicity, and star formation rate. Modeling the stellar component of galaxies is done by the default BC03 \citep{2003MNRAS.344.1000B} spectral evolution synthesis models with the \cite{2003PASP..115..763C} initial mass function. We considered a set of six discrete metallicities: 0.0001, 0.0004, 0.004, 0.008, 0.02, 0.05. A delayed SFH with optional exponential burst and thirteen different values of star formation timescale ranging from 1 to 12 Gyr is assumed. The modified \cite{2000ApJ...533..682C} attenuation law is used with a ratio of total to selective extinction for the extinction curve applied to emission lines $r_{\mathrm{V}} = $ 3.1 (the average value for the Milky Way). Dust emission model \citep{2014ApJ...780..172D}, nebular and rest-frame parameters modules are also used.

\subsection{Fitting and quality assessment}

\begin{figure}
\includegraphics[width=1.1\linewidth]{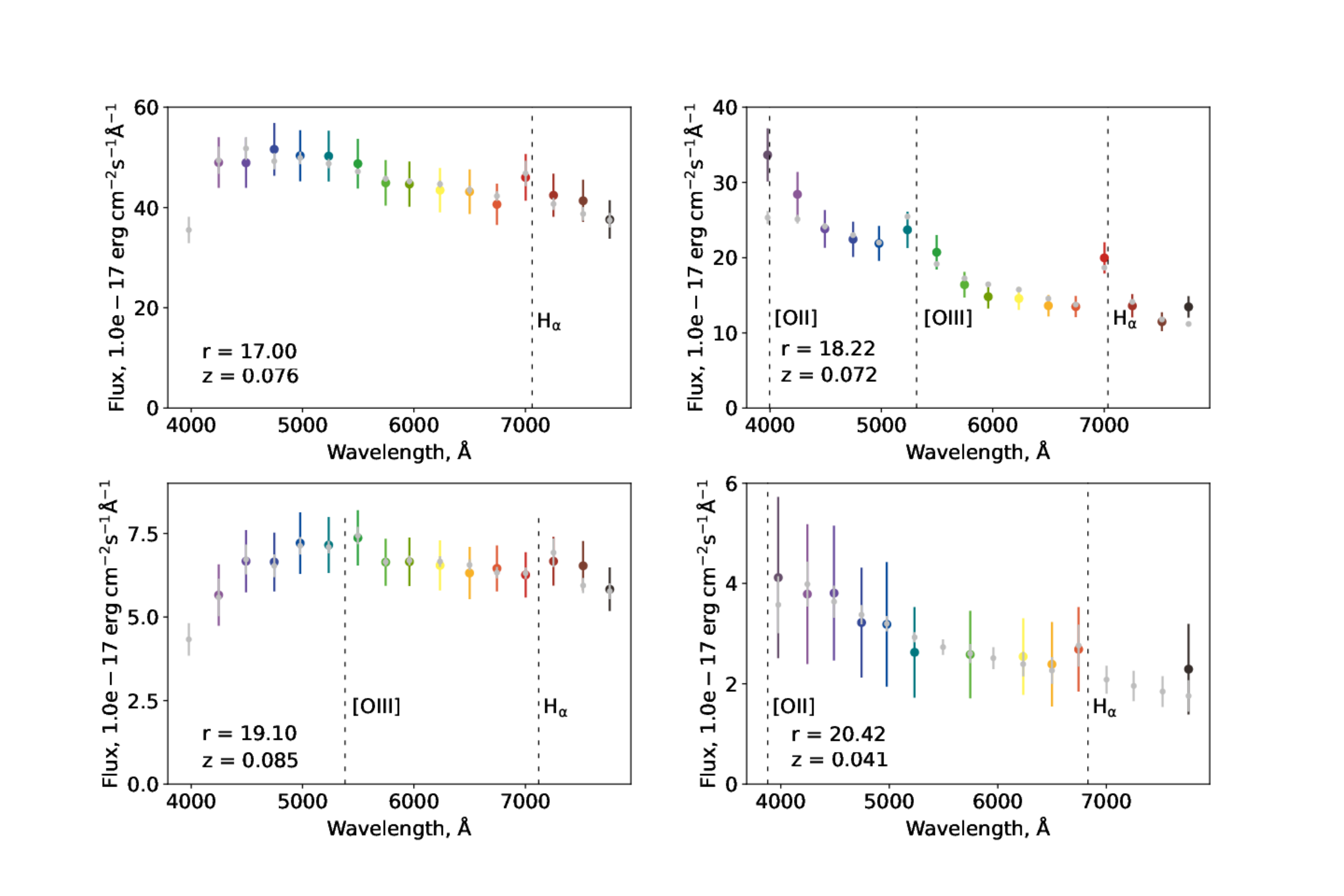}
\caption{MAG\_AUTO SEDs for different galaxies with brightness $17.0 \le r \mathrm{SDSS} \le 20.42$ mag in the redshift range $z = 0.04-0.08$ are shown by colour dots. The colour bars around these dots show the $\pm 1 \sigma$ variation. The Bayes model fitted to the spectral energy distribution with \textsc{cigale} is plotted as gray dots, and the gray bars show the magnitudes of the mean model at the $\pm 1\sigma$ uncertainty level.}
\label{ris:image8}
\end{figure}

SED-fitting reproduces well enough the values in 16 medium-band and $u$ SDSS filters of various types of galaxies in gMOSS sample within uncertainty and regardless of the redshift and brightness range. Figure \ref{ris:image8} shows nearby galaxies (up to $z=0.1$) in the magnitude range from 18.0 to 21.0 mag in the $r$ SDSS filter. The main spectral features of early (e.g. H+K Ca \textsc{ii}, Mg) and late-type galaxies (e.g. $\mathrm{H \; \alpha}$, [O \textsc{ii}] and [O \textsc{iii}]) are also well reproduced. SED-fitting uncertainties increase towards the fainter magnitudes because of the higher uncertainties in the data, which also increases for fainter objects.

\begin{figure}
\includegraphics[width=1.1\linewidth]{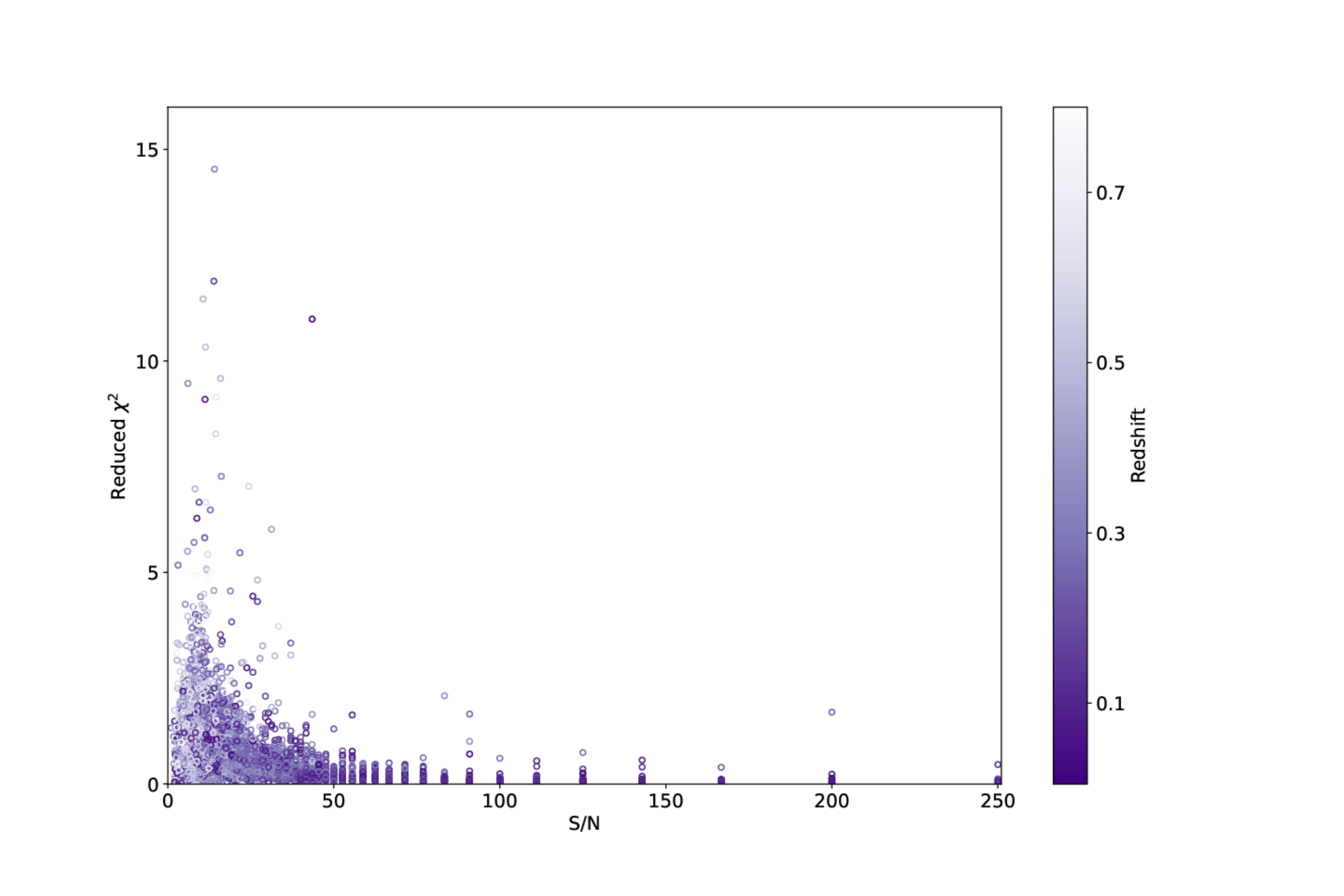}
\caption{Distribution of reduced $\chi^{2}$ parameter and signal-to-noise ratio (S/N) in $r$ SDSS filter for objects with MAG\_AUTO photometry. Colour bar shows the redshift of each galaxy. The discretization of the signal-to-noise ratio distribution for bright galaxies is because of the limitation of photometric accuracy.}
\label{ris:image9}
\end{figure}

The spectral energy distributions from the gMOSS catalogues are well reproduced by the SED-fitting code \textsc{cigale}. The quality of the correspondence for the entire sample of galaxies can be estimated using the reduced $\chi^{2}$-parameter. The dependence of the $\chi^{2}_{\mathrm{reduced}}$ parameter on the signal-to-noise ratio is shown in Figure \ref{ris:image9}, the colour bar corresponds to the redshift for each object. The discretization of the signal-to-noise ratio distribution for bright galaxies is due to the limitation of photometry accuracy. As expected, there is some inverse dependence for the chi-squared parameter value on the signal-to-noise ratio: $\chi^{2}_{\mathrm{reduced}}$ parameter estimates are minimal for bright objects, while with a decrease in the signal-to-noise ratio, the SED-fitting of objects becomes less accurate. The $\chi^{2}_{\mathrm{reduced}}$ parameter increases with a decrease in the signal-to-noise ratio and an increase in the redshift value. About half of the objects have $\chi^{2}_{\mathrm{reduced}} \le 0.5$, and for $\sim 60$ per cent of objects,  $\chi^{2}_{\mathrm{reduced}} \le 1.5$. We obtained rather small values for the reduced $\chi^{2}$ parameter, which shows a good accuracy in estimating the physical quantities of the galaxies in the sample.

We can fit 12,281 galaxies of the 16,509 objects from the full sample ($\sim \! 74$ per cent) with the \textsc{cigale} SED-fitting code. This number of objects suffices to study the statistical dependencies of the physical properties of galaxies in the entire studied redshift range $0< z \le 0.8$.

\subsection{Distributions of stellar populations properties}

\begin{figure*}
\includegraphics[width=1.0\linewidth]{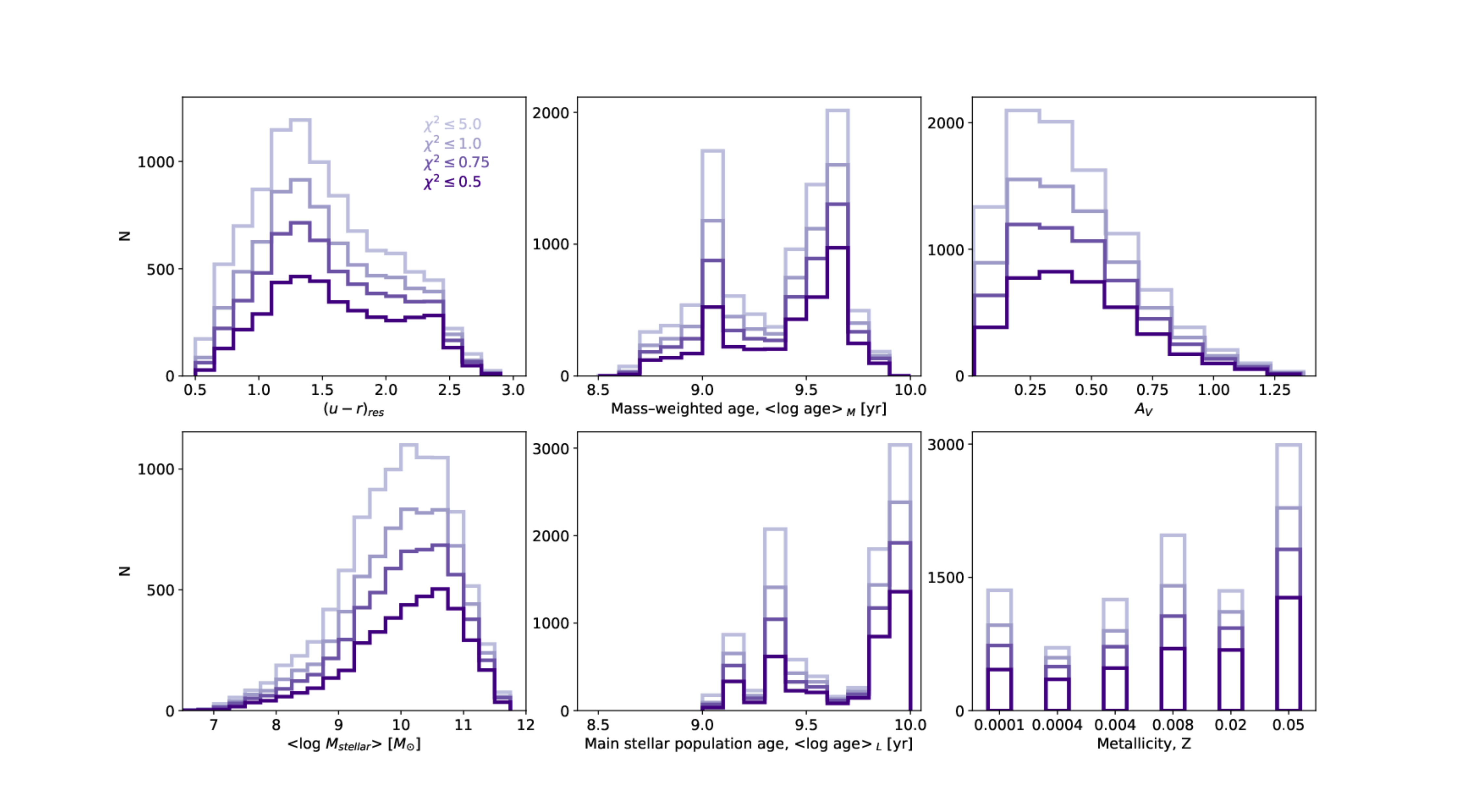}
\caption{Distributions of properties of stellar populations obtained with \textsc{cigale} by SED-fitting MAG\_AUTO magnitudes for the rest-frame $(u - r)_{\mathrm{res}}$ colour, the mass–weighted age, extinction $\mathrm{A_v}$, total stellar mass, age of the main stellar population in the galaxy and metallicity. The purple colour gradient shows different values of the reduced $\chi^{2}$ parameter.}
\label{ris:image10}
\end{figure*}

Fig. \ref{ris:image10} shows distributions of selected physical properties of fitted galaxies. The purple gradient lines show the quality of SED-fitting (reduced $\chi^{2}$-parameter). The colour $(u - r)_{\mathrm{res}}$ shows the bimodal distribution of galaxies, where the maximum density is at $(u - r)_{\mathrm{res}} \sim \! 1.5$ mag for blue galaxies and $(u - r)_{\mathrm{res}} \sim \! 2.5$ mag for the red sequence. The bimodal distribution becomes more clear for objects with a decrease in the value of the $\chi^{2}_{\mathrm{reduced}}$ parameter. 
The mass-weighted age distribution <log age>$_\mathrm{M}$ [yr] has two peaks on $\sim \! 9.0$ and $\sim \! 9.6$ dex.

The extinction $A_{\mathrm{V}}$ for the sample of galaxies is distributed in the range from 0 to 1.4 mag, which is slightly less than the ranges obtained in other works \citep[for example, in the][the extinction is in the range from 0 to 2.0 mag]{2021arXiv210213121G}. This is because of the choice of the observation field, which is in an area with a very low density of neutral hydrogen on the visual beam <$N_{\mathrm{H}}\mathrm{>}=10^{20} \; \mathrm{cm^{-2}}$, which is not much higher than the absorption value in the "Lockman Hole" region \citep{1986ApJ...302..432L}, where the lowest absorption on the visual beam is observed for the northern sky <$N_{\mathrm{H}}\mathrm{>}=4.5 \cdot 10^{19} \; \mathrm{cm^{-2}}$.

The stellar mass distribution ranges from $\sim \! 7$ to $\sim \! 11.5$ dex. 
The age of the main stellar population distribution <log age> [yr] has two peaks on $\sim \! 9.4$ and $\sim \! 9.9$ dex.
Metallicity is distributed discretely due to the initial conditions for calculating models.

\subsection{Mass–colour diagram}

\begin{figure*}
\includegraphics[width=1.1\linewidth]{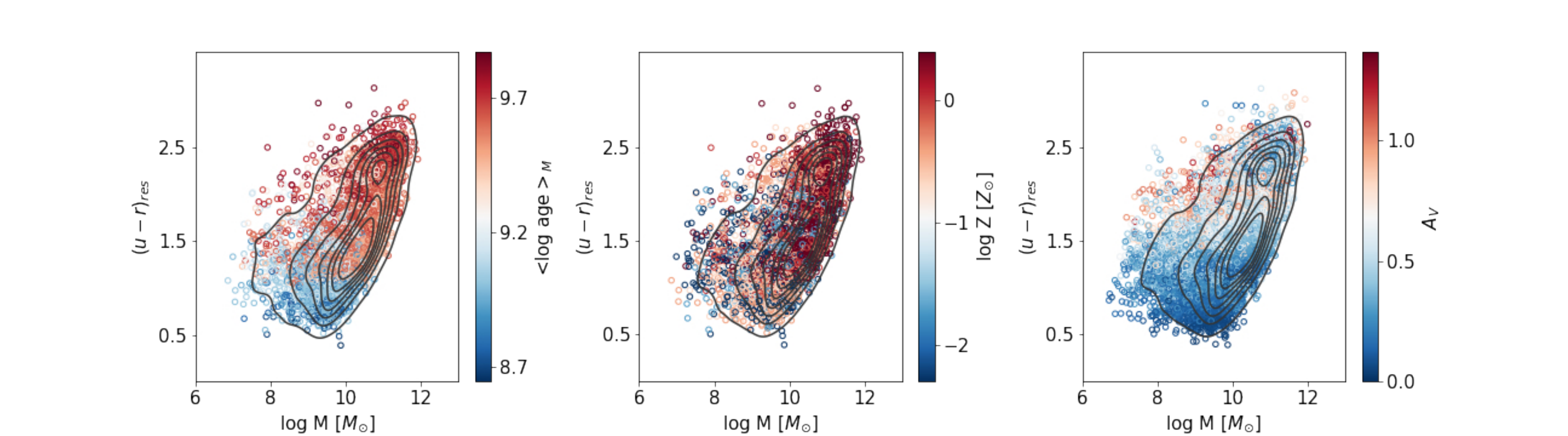}
\caption{Mass–colour relation, with the coloured bar showing the stellar population properties of the galaxies: age, metallicity and extinction.
Properties were derived using the MAG\_AUTO photometry.}
\label{ris:image11}
\end{figure*}

To study the differences between the red and blue populations of galaxies, it is customary to use mass-colour diagrams \citep{2019A&A...631A.156D, 2014MNRAS.440..889S, 2013A&A...558A..61M}. The rest-frame colour $(u - r)_{\mathrm{res}}$ shows that the galaxies are quite clearly divided into these two populations. Fig. \ref{ris:image11} shows the mass-colour diagrams in combination with the parameters of the mass-weighted age, metallicity, and extinction. The mass-colour diagram with the mass-weighted age parameter (Fig. \ref{ris:image11}, \emph{left}) shows the apparent separation of galaxies into younger blue galaxies with
$\mathrm{ log(age)}_\mathrm{M}$ [yr] down to $\sim 9.4$ dex and older red ones  with
$\mathrm{ log(age)}_\mathrm{M}$ [yr] above $\sim 9.5$ dex. Most galaxies have solar metallicity, however, super-solar values are often found for red sequence galaxies and low-metallicity values for blue galaxies. The extinction value practically does not depend on which population of galaxies it is determined for. However, $A_{\mathrm{V}}>1.0$ mag is typical for galaxies with $(u - r)_{\mathrm{res}} \sim 1.5$ mag, which roughly corresponds to the galaxies from the green valley. This is consistent with the assumption that the green valley galaxies are star-forming galaxies from the blue cloud \citep{2019A&A...631A.156D, 2021arXiv210213121G}.

\subsection{The accuracy of determining the properties of the stellar populations}

Uncertainties in determining the physical properties of galaxies were obtained within the SED-fitting process by the \textsc{cigale} code. Figure \ref{ris:image13} shows the distribution of the accuracy of the physical properties of galaxies in the colour-mass diagram. The parameter of the age of the main stellar population of the galaxy has the best precision of estimation: $\sigma(\mathrm{ log(age)}_\mathrm{L}) = 0.05 \pm 0.04$ dex. Estimates of the accuracy of total stellar mass, the mass–weighted age and  the rest-frame $(u - r)_{\mathrm{res}}$ colour, are fairly well obtained: $\sigma(\mathrm{log}(M)_{[M_\odot]}) = 0.12 \pm 0.05$ dex, $\sigma(\mathrm{ log(age)}_\mathrm{M}) = 0.10 \pm 0.04$ dex and $\sigma(u - r)_{\mathrm{res}} = 0.19 \pm 0.08$ mag. The stellar extinction  precision is $\sigma(A_{\mathrm{V}}) = 0.28 \pm 0.09$ mag. The accuracy depends on the signal-to-noise ratio and resolution in SEDs. The accuracy estimates obtained are quite similar to the accuracy estimates of physical parameters from other works, for example from \citep{2021arXiv210213121G}. The average value of S/N in our sample is $\sim 10$ but our resolution is over two times worse than in miniJPAS survey (FWHM filters is 100 \AA) thus, the accuracy of our estimates is lower.

\begin{figure*}
\includegraphics[width=1.0\linewidth]{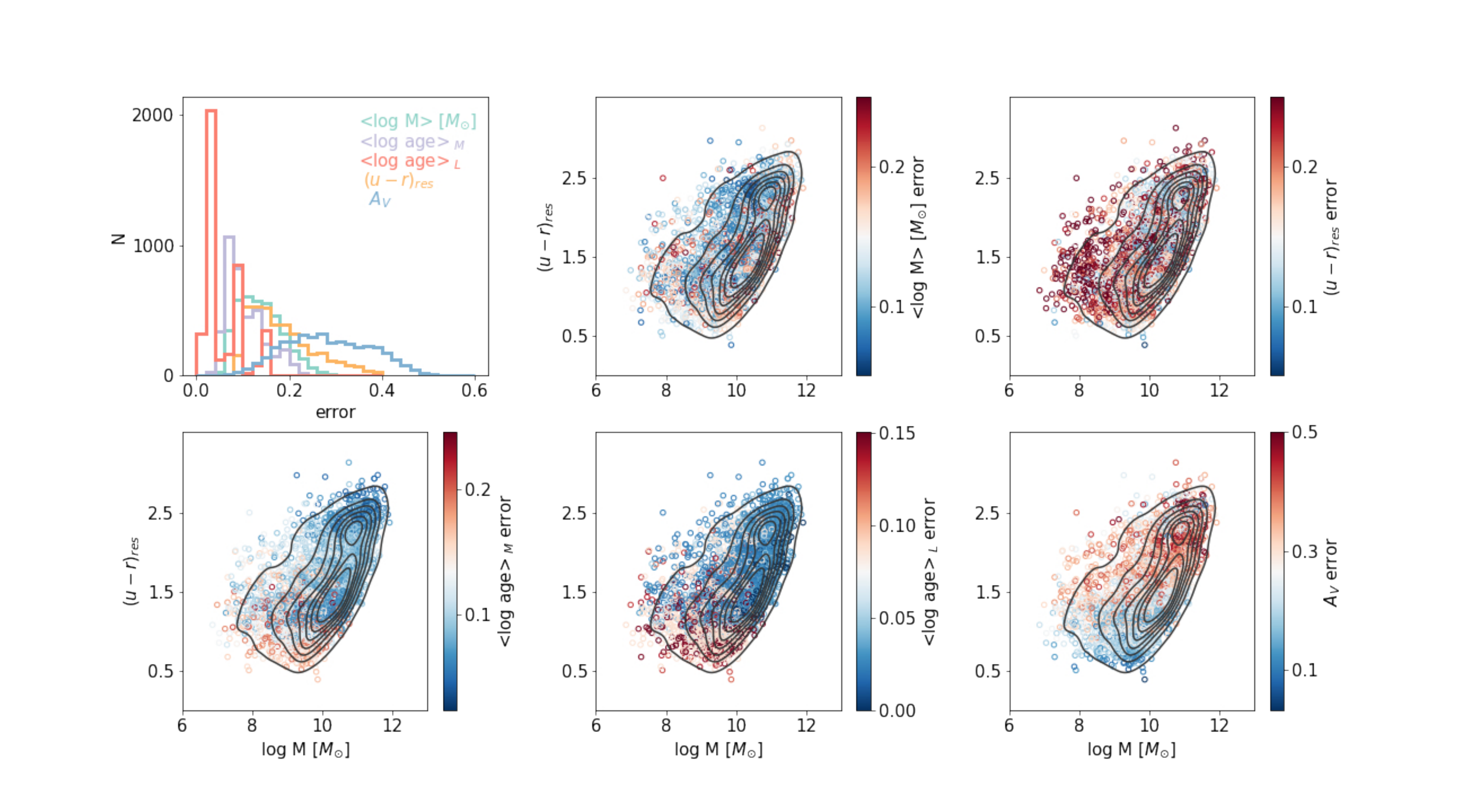}
\caption{Distribution of standard deviations of galaxy properties (\emph{top-left panel}) and the colour–mass plane
coloured by the standard deviation of the stellar population properties: total stellar mass, the rest-frame $(u - r)_{\mathrm{res}}$ colour, the mass–weighted age, age of the main stellar population in the galaxy and V-band attenuation $\mathrm{A_v}$.}
\label{ris:image12}
\end{figure*}

The colour-mass diagrams in Figure \ref{ris:image12} show that the value of errors in determining the physical parameters of galaxies is not the same for different populations of galaxies. The bimodality of the distribution is clearly visible in the diagrams showing the error of determining both the mass-weighted age and the age of the main stellar population. For red sequence galaxies, this value rarely exceeds ($\sigma(\mathrm{ log(age)}_\mathrm{M}) \le 0.1$ dex and $\sigma(\mathrm{ log(age)}_\mathrm{L}) \le 0.1$ dex). 
In addition, the mass of the main stellar population is also determined better for red galaxies $\sigma(\mathrm{log}(M)_{[M_\odot]})\le 0.1$ dex than for blue cloud galaxies and $\sigma(u - r)_{\mathrm{res}}$ is constrained slightly better in the red sequence than in the blue cloud. The uncertainties in $A_{\mathrm{V}}$ are the same for blue cloud galaxies, while a higher value of $\sigma(A_{\mathrm{V}})$ corresponds to more massive galaxies in the red sequence. 

\section{Discussion}
\label{s:Discussion}

The accuracy of the estimation of photometric redshifts in our sample of galaxies ($\sigma_\mathrm{{NMAD}} < 0.0043$) makes it possible to study the evolution of the physical properties of galaxies with cosmic time. Our survey is actually the largest in the area (the total homogeneous area is 2.386 $\mathrm{deg^2}$) among the mid-band deep surveys at the moment, which makes its data fairly homogeneous and complete. Sufficient SED resolution (FWHM of a medium-band filter is 250 \AA) makes it possible to divide galaxies into a red sequence and a blue cloud and to study their properties independently of each other through cosmic time.

\begin{figure*}
\includegraphics[width=1.0\linewidth]{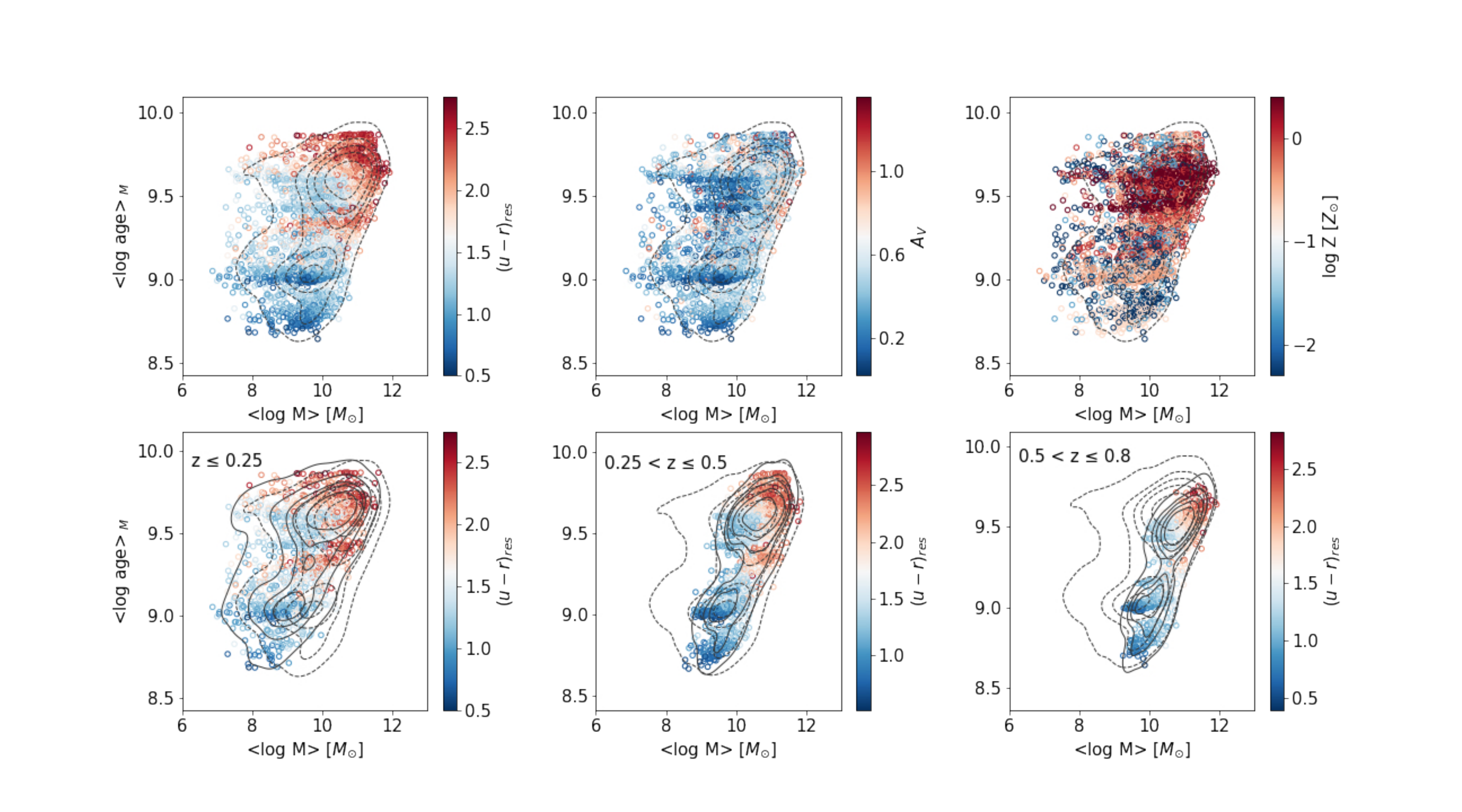}
\caption{\emph{The top panel}: Distributions of the rest-frame $(u - r)_{\mathrm{res}}$ colour, V-band attenuation $A_{\mathrm{V}}$, and metallicity across the mass–age relation. The values of all the parameters are coded according to the colour bars. \emph{The bottom panel} illustrates the points and contour distribution of the rest-frame $(u - r)_{\mathrm{res}}$ colour in the mass–age relation at $z \le 0.25$, $0.25 < z \le 0.5$, and $0.5 < z \le 0.8$. The dashed contours show the distribution of galaxies with $z \le 0.8$. The discretization of the age values of galaxies occurs because of the limitations of computing power for the initial set of calculated values.}
\label{ris:image13}
\end{figure*}

In our work, there was no goal of a detailed study of the evolution and formation of galaxies depending on the redshift. We investigated the physical properties of galaxies in general and compared the results obtained with previous works by other authors. For the correct interpretation of the results, it is necessary to consider that there is a limit on the visibility of galaxies of a certain absolute galaxy magnitude at a certain redshift \citep[Malmquist bias, ][]{1922MeLuF.100....1M}. Thus, the least massive galaxies fall out of the field of our consideration with an increase in redshift. We leave the study of the bias to the region of more massive galaxies outside this work. However, it should be noted that Figure \ref{ris:image13} shows that galaxies with a mass less than $10^8 {\mathrm{M_\odot}}$ occur only at lower redshifts.

To analyze the general physical properties of galaxies, we planned to use photometry obtained in Kron-like apertures. However, in the study's course, data were obtained that showed that Kron-like photometry is effective only for galaxies up to $r_{\mathrm{AB}} = 20.5$ mag, then the uncertainties associated with determining the Kron radius for objects arise due to the insufficient signal-to-noise ratio. Therefore, for objects fainter than $r_{\mathrm{AB}} = 20.5$ mag, we used aperture photometry.

\subsection{Evolution of galaxy populations in the mass - age diagram}

Like the colour-mass diagram, the bimodality in the distribution of galaxies is quite clearly visible in the mass-age $\mathrm{log}(M)\!_{[\mathrm{M_\odot}]} -\mathrm{ log(age)}_\mathrm{M}$ (Fig. \ref{ris:image14}). The oldest galaxies in the sample have redder colours, a larger mass, and a higher metallicity value.

In recent works \citep{2003MNRAS.341...54K, 2014A&A...562A..47G, 2021arXiv210213121G}, it is shown that bimodality is also observed because galaxies from the blue cloud have a linear dependence of age on the stellar mass of the galaxy. In our work, we also find this dependence (Fig. \ref{ris:image13}, \emph{bottom panels}). This dependence is most clearly manifested if we analyze galaxies from the same cosmic epochs. For this purpose, we divided our sample into three bins: $z \le 0.25$, $0.25 < z \le 0.5$ and $0.5 < z \le 0.8$. At a small redshift, the mass-age ratio shows a change in the dependence's slope for galaxies with a mass greater than $\mathrm{log}(M)\!_{[\mathrm{M_\odot}]} = 10.5$ dex, for which the ratio $\mathrm{log}(M)\!_{[\mathrm{M_\odot}]} -\mathrm{ log(age)}_\mathrm{M}$ is becoming slightly sloping in comparison with the late-type galaxies. This mass limit refers to the green valley galaxies, which are the transition from the blue cloud to the red sequence. Analyzing the red galaxies of all three bins by redshift, it is noticeable that with decreasing age, the mass of the galaxy also decreases. Thus, we can assume that the least massive galaxies formed in later cosmic epochs.

\subsection{Evolution of galaxy populations in the mass - age diagram with extinction correction}

Dust extinction reddens galaxies, so it is important to study the colour - mass ratio not only for the colour $(u - r)_{\mathrm{res}}$ in the rest-frame but also for the colour $(u - r)_{\mathrm{int}}$ in the rest-frame corrected for extinction (so-called intrinsic colour). After extinction correction, the ratio between the galaxies of the red sequence and the blue cloud changes significantly \citep{2014MNRAS.440..889S, 2019A&A...631A.156D}.

\begin{figure*}
\includegraphics[width=1.1\linewidth]{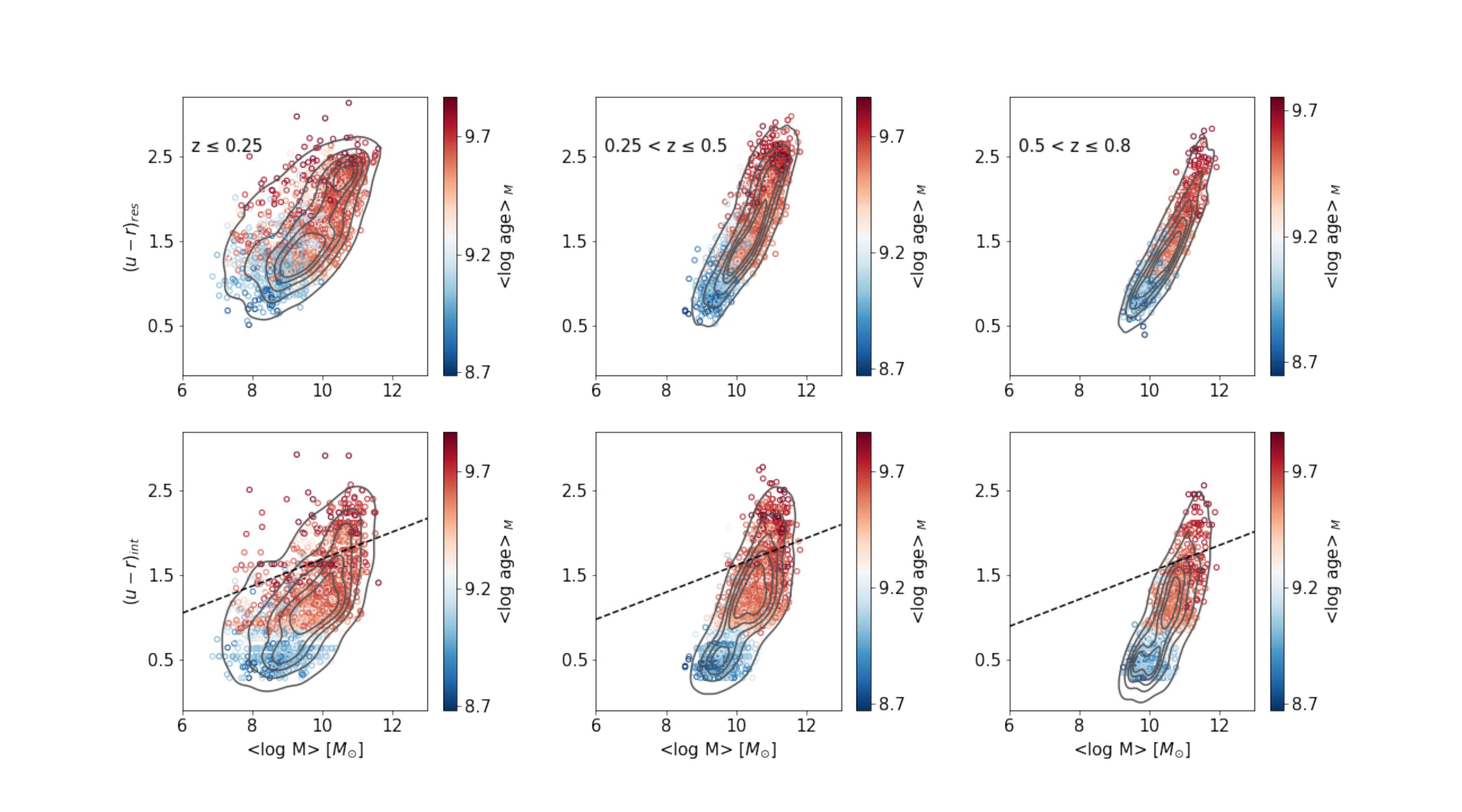}
\caption{Colour-mass relation for rest frame (top panels) and intrinsic (bottom panels) colour (u - r) in the three redshift bins: $z \le 0.25$, $0.25 < z \le 0.5$, and $0.5 < z \le 0.8$ (from left to right). The dashed line shows the limiting intrinsic colour for galaxy classification for the mean redshift in each bin (see more details in Section \ref{ss:blue.red.separation.}).}
\label{ris:image14}
\end{figure*}

Figure \ref{ris:image14} shows the distribution of galaxies in the colour-mass diagram for the rest-frame colour $(u - r)_{\mathrm{res}}$  and for the intrinsic colour $(u - r)_{\mathrm{int}}$ in three redshift bins: $z \le 0.25$, $0.25 < z \le 0.5$, and $0.5 < z \le 0.8$. It can be seen that the distribution of galaxies in the colour - mass diagram for the rest frame colour $(u - r)_{\mathrm{res}}$ and the intrinsic colour $(u - r)_{\mathrm{int}}$ differ significantly in each redshift bin. With introducing of the extinction correction, a significant part of the galaxies from the green valley passes into the blue cloud. This is because a significant part of the green valley galaxies is obscured star-forming galaxies \citep[30–65 per cent;][]{2019A&A...631A.156D}, the proportion of which depends on the redshift and stellar mass. The proportion of such galaxies in the near Universe is not large, but it is not negligible \citep{2014MNRAS.440..889S}. At high redshifts, the proportion of dusty green valley galaxies increases even more.

\subsection{Identification of blue and red galaxies}
\label{ss:blue.red.separation.}

We use the technique of dividing galaxies into a red sequence and a blue cloud, developed in the \citep{2019A&A...631A.156D} for galaxies from the ALHAMBRA survey in the redshift range from $0.1 \le z \le 1.1$. The ALHAMBRA survey used a set of 20 mid-band filters in the optical range, as well as near-infrared filters J, H, and $\mathrm{K_s}$. Using the SED-fitting method, the authors determined the physical properties of galaxy populations, such as stellar mass, rest-frame colour, and extinction of each galaxy. This allowed them to divide the galaxies from the sample into a red sequence and a blue cloud. They found the proportion of dusty star-forming galaxies in the green valley using intrinsic colours and contamination in the samples of rest-frame galaxies, determined using classical colour diagrams, caused by the obscuring of star-forming galaxies. The authors concluded that the use of a mass - colour diagram can reduce the contamination of a part of the galaxies from the red sequence by 20 per cent compared with the use of colour - colour diagrams.

In the \citep[][eq.3]{2019A&A...631A.156D}, the equation defining the limit of the separation of galaxies into a red sequence and a blue cloud used the true colours for the $m_{F365}$ and $m_{F551}$ filters. In \citep[][ eq.5]{2021arXiv210213121G}, this criterion was recalculated into colours $(u - r)_{\mathrm{int}}$ that similar to our photometric system. So we can use this equation \ref{eq:blue_red_sep} with no corrections.

\begin{equation}
    (u-r)_\mathrm{res}^\mathrm{lim} = 0.16 \cdot (\mathrm{log} M - 10.0) - 0.3 \cdot (z-0.1) + 1.7,
	\label{eq:blue_red_sep}
\end{equation}

where $z$ is the photometric redshift of each
galaxy and $\mathrm{log} M$ is its stellar mass. 

Galaxies from gMOSS survey are labeled as quiescent if intrinsic colour is redder than the limiting value $(u-r)_\mathrm{res}^\mathrm{lim}$ else galaxies are star-forming.

\subsection{Characterization of blue and red galaxies}

Using the criterion for selecting star-forming and quiet galaxies from the equation \ref{eq:blue_red_sep}, we obtained that about $\sim 86$ per cent of the sample from our survey are blue cloud galaxies, and the share of red sequence galaxies accounts for the remaining $\sim 14$ per cent. 



\begin{figure*}
\includegraphics[width=1.0\linewidth]{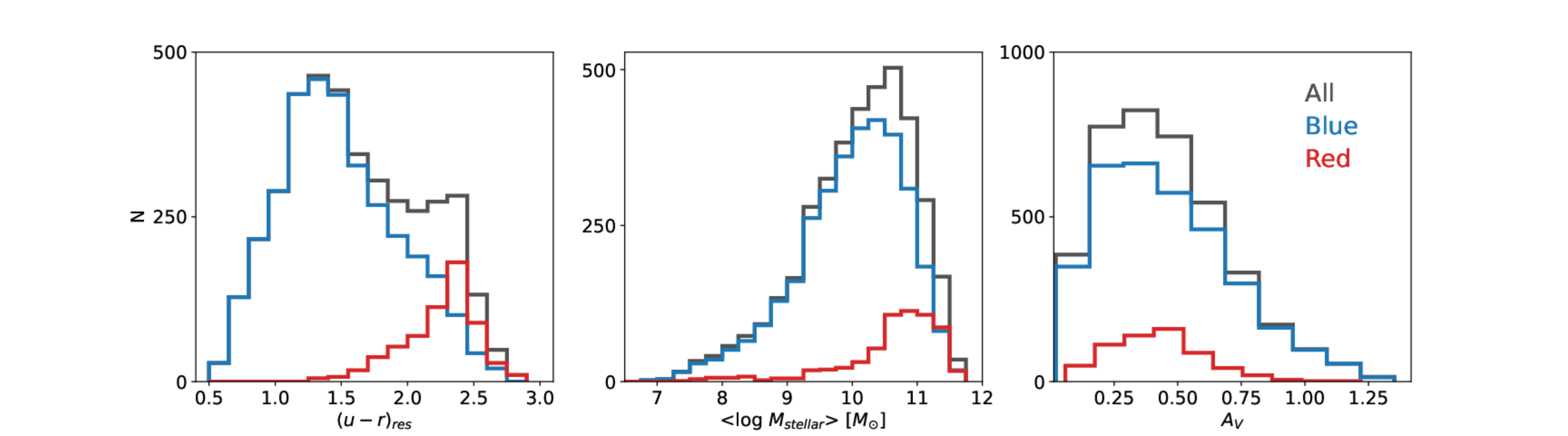}
\caption{Distributions of properties of stellar populations for two separate galaxy populations for the rest frame (u-r) colour, total stellar mass, extinction $A_v$. The complete sample is indicated by a black line, the galaxies of the red sequence are red, the blue clouds are blue.}
\label{ris:image16}
\end{figure*}

Figure \ref{ris:image16} shows distributions of properties of stellar populations for two separate galaxy populations for the rest-frame $(u-r)_\mathrm{res}$ colour, total stellar mass, extinction $A_v$. In common red galaxies have a redder colour $(u-r)_\mathrm{int}$ and lower extinction. Also, galaxies from the red sequence are more massive.

The separation of galaxies into two groups using the colour - mass diagram can be justified either by differences in the evolutionary path of the galaxy or by differences in stellar content. Next, we will discuss the dependence of the properties of stellar populations of galaxies on redshift.

\begin{figure*}
\includegraphics[width=1.0\linewidth]{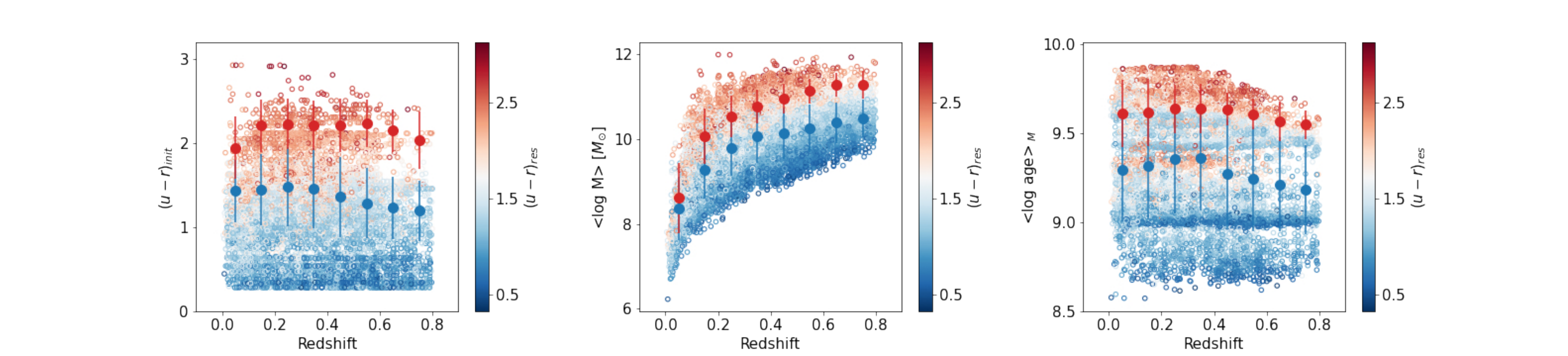}
\caption{Evolution of the intrinsic colour, stellar mass, and age of galaxies. Dots represent
the average values of each property in each redshift bin. The dispersion with respect to the average values is shown as error bars. Blue and red dots correspond to the blue cloud and red sequence galaxies, respectively.}
\label{ris:image17}
\end{figure*}

In Figure \ref{ris:image17}, we investigate the evolution of the intrinsic colour, stellar mass, and age of galaxies through redshift. We obtain average estimates of these characteristics of galaxies for each bin by redshift. Red and blue galaxies are distributed on these diagrams according to their stellar component.

The $(u-r)_\mathrm{int}$ colour distribution shows that both red and blue galaxies become bluer at high redshifts. For red galaxies, the first redshift bin shows a slightly bluer  $(u-r)_\mathrm{int}$ colour than on intermediate redshifts.

On average, blue galaxies are about 0.8 dex lighter than red ones. At the same time, a significant decrease in this difference is observed for the galaxy of the near Universe ($z \le 0.1$), this may be due to a shift caused by the fact that faint galaxies become invisible with an increase in redshift. In addition, the average mass in each of the galaxy populations increases with the growth of the redshift should be associated with the same reason.

The galaxies of the red sequence and the blue cloud are quite well separated in the diagram showing the stellar age. At any redshift, red galaxies are older by $\sim 0.35$ dex. This is probably a consequence of different histories of star formation (SFH) and/or the formation of blue and red galaxies. The age of both blue and red galaxies decreases with increasing redshift, indicating ongoing star formation and/or reflecting a biased sample for low-mass galaxies at higher redshifts.

\subsection{Star formation rate density}
\label{SFRD}

The most prominent result in the study of the evolution of galaxies through the redshift is that it has been established that the star formation rate density has a peak at $z \sim 3$ and is decreasing until the present cosmic time \citep{1996ApJ...460L...1L, 1998ApJ...498..106M, 2006ApJ...651..142H, 2007MNRAS.379..985F, 2013MNRAS.433.2764G, 2014ARA&A..52..415M, 2018MNRAS.475.2891D}.

In order to study the star formation rate density (SFRD), it is necessary to study the incompleteness of our sample and the limits of detection of the stellar mass depending on the redshift. It is necessary to find the minimum and maximum redshifts ($z_\mathrm{min}$ and $z_\mathrm{max}$) at which each galaxy of the sample can be detected because of the detection limits of cMOSS. The sample consists of galaxies in the range from 15.1 to 22.5 mag in $r$ SDSS filter, which we used as constraints. In order to estimate the maximum and minimum redshifts, we used the limits for detecting stellar magnitudes in cMOSS sample and the properties of the stellar population obtained using the \textsc{cigale} code. The average values of $z_\mathrm{min}$ and $z_\mathrm{max}$ were obtained in bins by the stellar mass with a width of $\Delta \mathrm{log}(M) = 0.2$ dex.

\begin{figure}
\includegraphics[width=1.0\linewidth]{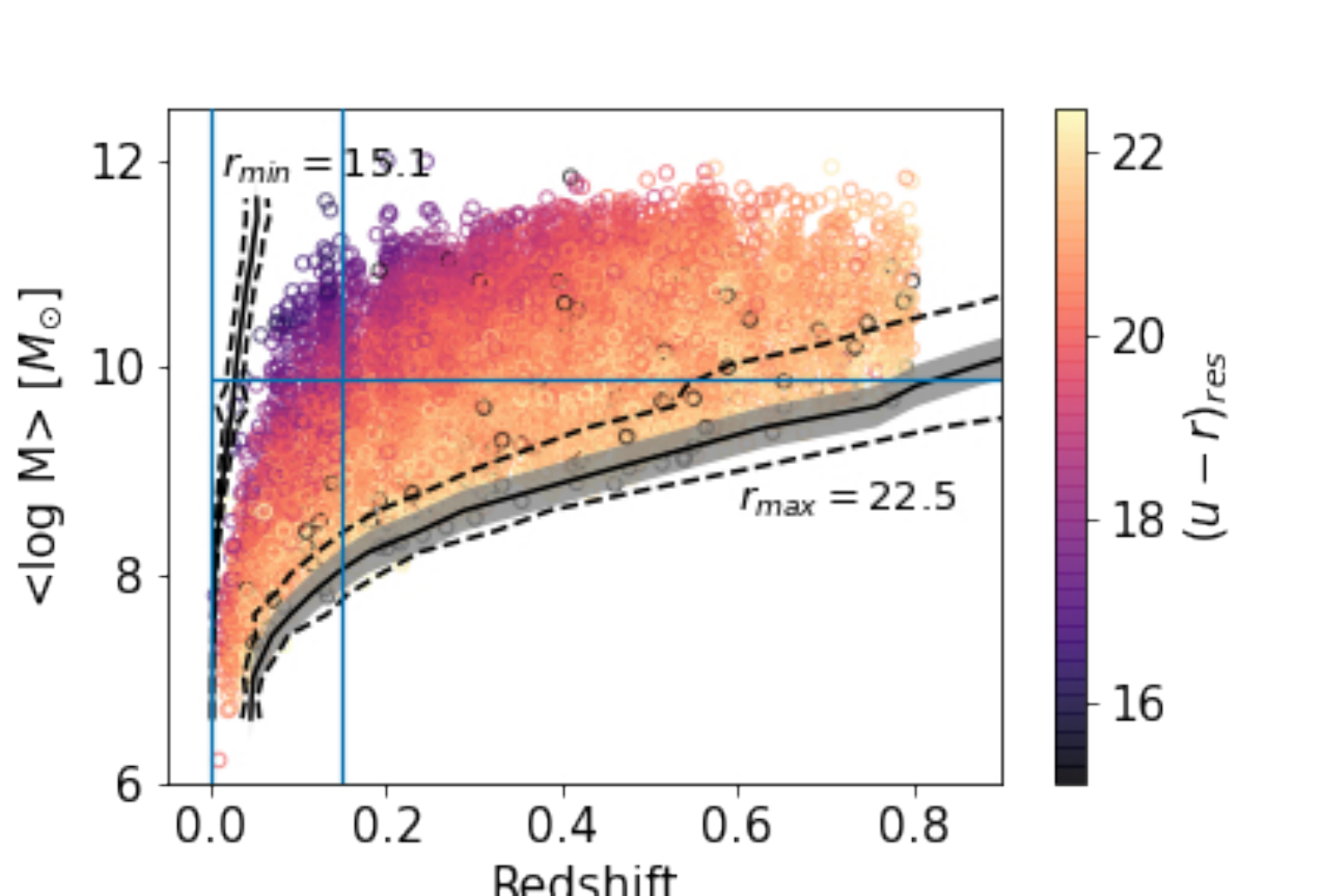}
\caption{Distribution of redshifts and stellar masses obtained by \textsc{cigale} for each of the galaxies in the sample. The
black lines show the $z_\mathrm{max}$ and $z_\mathrm{min}$ values that correspond to the limiting magnitudes from 15.1 to 22.5 mag of the cMOSS galaxy survey. The shaded regions illustrate the dispersion of limiting magnitude in the y-axis, and the dashed lines the dispersion in the x-axis. The blue lines show the sub-sample of galaxies with $z \le 0.15$ used to explore the evolution of the star formation rate density. Colour bar code galaxy magnitudes in the $r$ SDSS band.}
\label{ris:image18}
\end{figure}

Figure \ref{ris:image18} shows the result of our estimates of the detection limit by stellar mass as a function of the redshift. Galaxies with a mass of $\mathrm{log}(M)\!_{[\mathrm{M_\odot}]} \sim 10$ dex can be detected in the entire redshift range ($z \le 0.8$) presented in the sample. Low-mass galaxies of $\mathrm{log}(M)\!_{[\mathrm{M_\odot}]} \sim 8.0$ can be detected up to $z = 0.15$. Also Figure \ref{ris:image18} show that we are able to study samples of galaxies with stellar masses above $\mathrm{log}(M)\!_{[\mathrm{M_\odot}]} \sim 8.9$, 9.5, and 9.9 dex at $z = 0.4$, 0.6, and 0.8, respectively.

Thus, our results allow us to study the SFRD for galaxies in the redshift range $z \le 0.15$ using the star formation history (SFH) for galaxies in the gMOSS sample obtained by the \textsc{cigale} code. At the upper limit of the redshift range ($z = 0.15$), the sample of galaxies includes galaxies with a mass above $10^8 \; \mathrm{M_\odot}$. Galaxies with a mass below this limit have contributed significantly to the history of star formation over the past 4 Gyr. In the cMOSS galaxy sample, such galaxies are detected in the range up to $z=0.15$, however, their number does not assess the incompleteness of the sample.

To get SFRD, considering the effect of the incompleteness of volume, we divided the star-formation rate (SFR) of each galaxy at $0.005 \le z \le 0.15$ by its maximum co-moving volume ($V_\mathrm{max}$). Since only a small number of galaxies with a mass below $10^8 \; \mathrm{M_\odot}$ are not observed at $z=0.15$ ($\sim 3$ per cent), we can assume that $V_\mathrm{max}$ is equal to the co-moving volume $V_\mathrm{c}$ at this redshift range: 

\begin{equation}
    V_\mathrm{max} = \Delta V = V_\mathrm{c}(z=0.15) - V_\mathrm{c}(z=0.005).
	\label{eq:blue_red_sep}
\end{equation}

However, the redshift bin $0.005 \le z \le 0.15$ includes a small number of galaxies from the total sample (1250 galaxies), so it is also necessary to estimate the cosmic variance for this redshift range in survey area (2.386 $\mathrm{deg^2}$). We used the equations for cosmic variance obtained in the work by \cite{2010MNRAS.407.2131D} and got the cosmic variance estimation 39.3 per cent for our survey parameters. This value is quite large and may explain the differences between SFRD values estimated for the same redshift interval in different articles. 

The SFRs were obtained from the parametric delayed SFH with optional exponential burs for a 1 Gyr resampling:
\begin{equation}
    \mathrm{SFR}\propto \frac{t}{\tau^2} \cdot \mathrm{exp}(-\frac{t}{\tau} ),
	\label{eq:delayedSFH}
\end{equation}
where $0 \le t \le t_0$ with $t_0$ the age of the onset of star formation, and $\tau$ the time at which the SFR peaks. Such a functional form providing a nearly linear increase of the SFR from the onset of star formation and after peaking at $t = \tau$, it smoothly decreases.

\begin{figure*}
\includegraphics[width=1.0\linewidth]{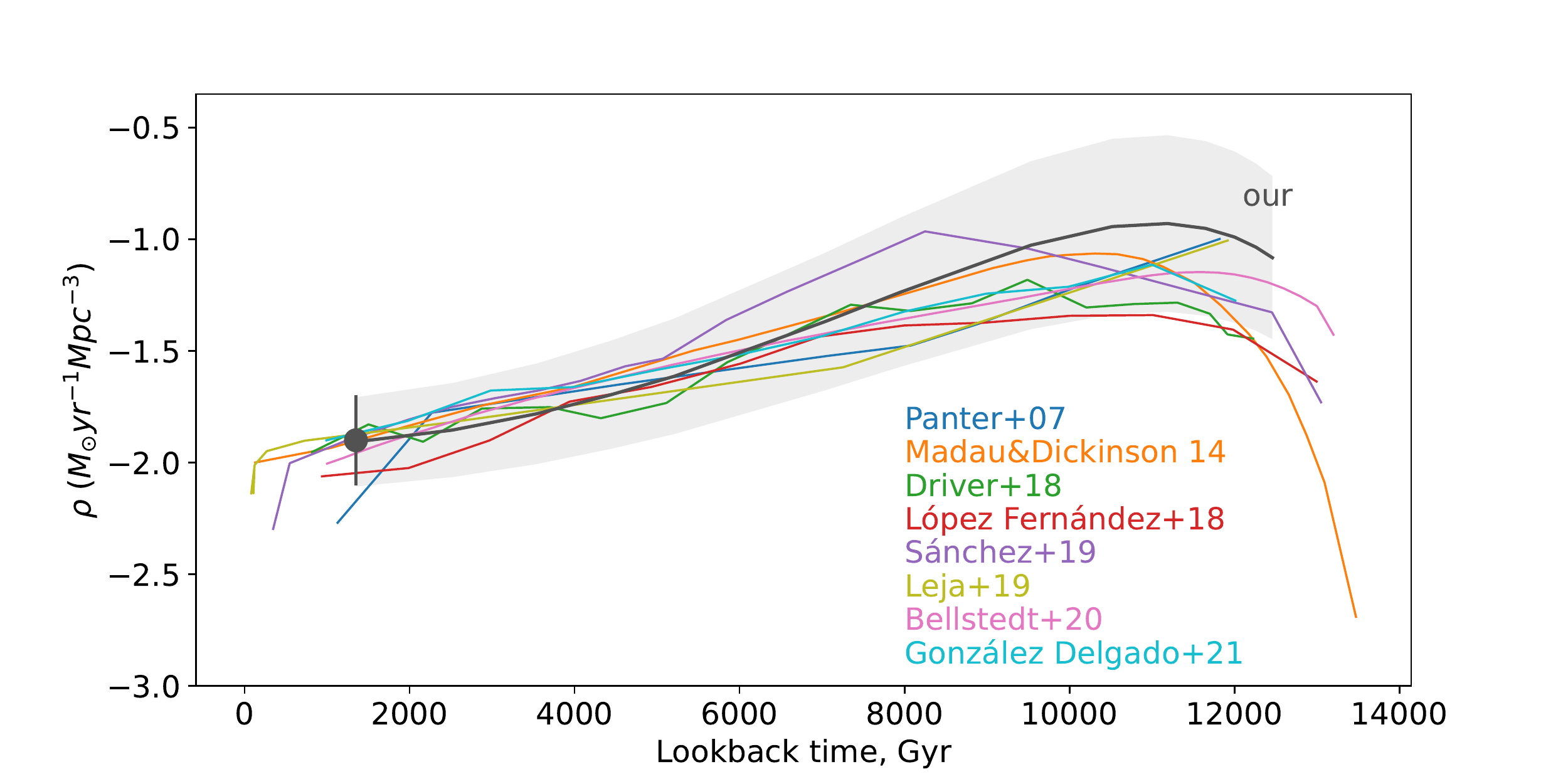}
\caption{SFRD evolution through cosmic time was obtained from the SED-fitting results (gray line) with the nearby galaxies ($0.05 \le z \le 0.15$, gray dot). The grey shadow area represents the uncertainties of the result. The different colour lines show the SFRDs obtained in other works.}
\label{ris:image19}
\end{figure*}

The error of log $\rho$ in each epoch is obtained by propagating the SFRD dispersion in each bin. The results shown in Figure \ref{ris:image19} demonstrate that log $\rho$ increases with redshift, up to $z \sim 3.0$ and then decreases.

We compared the results obtained with those already known in the literature in the works \citep{2007MNRAS.378.1550P, 2014ARA&A..52..415M, 2018MNRAS.475.2891D, 2018A&A...615A..27L, 2019MNRAS.482.1557S, 2019ApJ...876....3L, 2020MNRAS.498.5581B, 2021arXiv210213121G}. These results are obtained from different data samples and using different analysis techniques. Overall, we conclude that the analysis of gMOSS sample provides results that are in good agreement with cosmological surveys \citep{2014ARA&A..52..415M, 2018MNRAS.475.2891D} and the nearby galaxies analysis f SDSS \citep{2007MNRAS.378.1550P}, IFS CALIFA \citep{2018A&A...615A..27L}, and GAMA \citep{2020MNRAS.498.5581B}.

\subsection{Stellar-mass density}

Given the volume incompleteness effect from Section \ref{SFRD}, we can calculate the SMD as the ratio of the sum of the stellar mass of each galaxy obtained using the \textsc{cigale} code to the co-moving volume ($V_\mathrm{max}$), in the redshift bin $0.05\le z \le 0.15$. For other redshift ranges, the SMD value is calculated by integrating the SFRD function over cosmic time.

\begin{figure}
\includegraphics[width=1.0\linewidth]{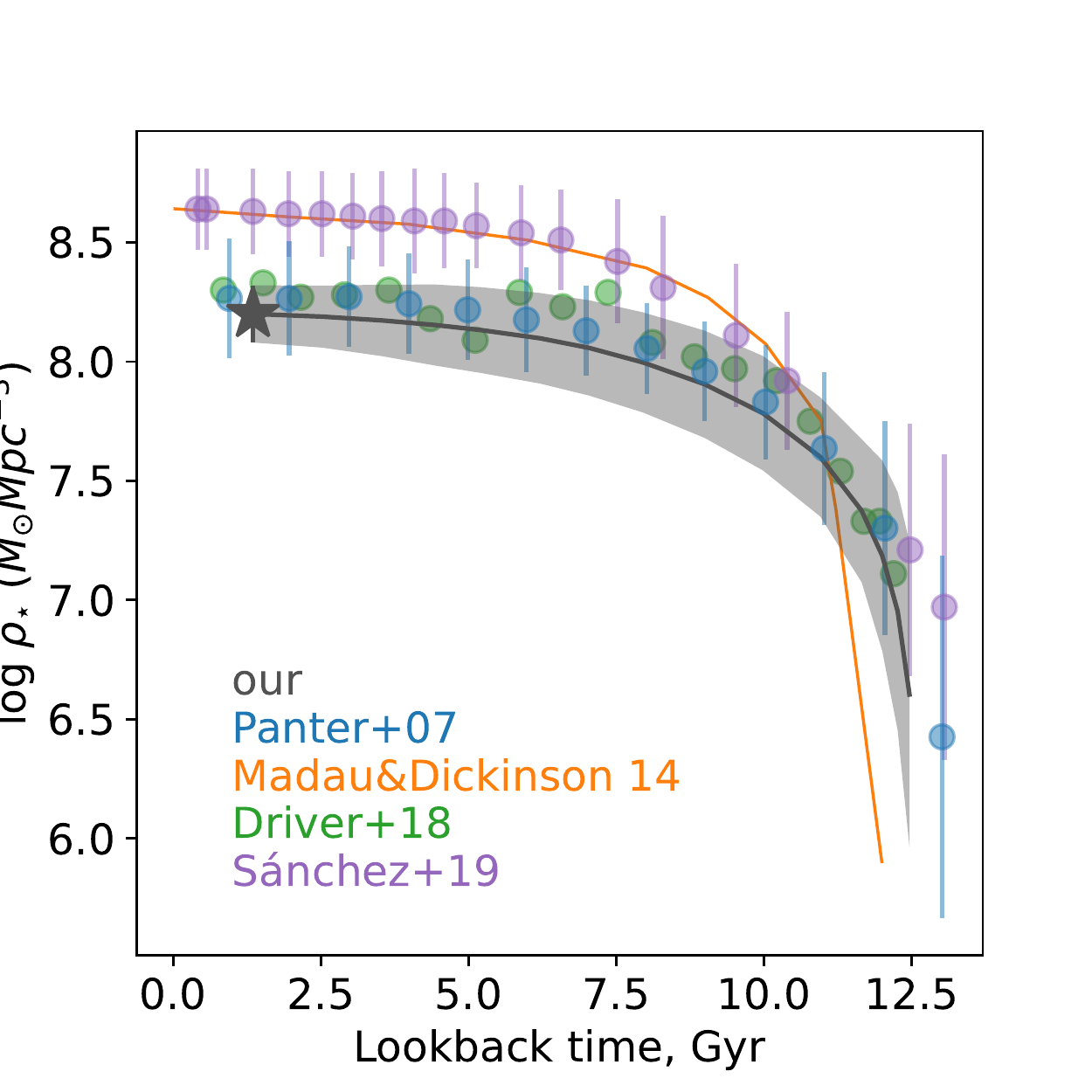}
\caption{SMD evolution through cosmic time was obtained from the SED-fitting results (gray line) with the nearby galaxies ($0.05 \le z \le 0.15$, star) and integration of SFRD function. The grey shadow area correspond to the uncertainties of the result. The different colour lines and markers show the SMDs obtained in other works.}
\label{ris:image20}
\end{figure}

Figure \ref{ris:image20} shows the SMD (stellar-mass density) as a function of cosmic time, compared to the \citet[][blue dots]{2014ARA&A..52..415M} fit, the \citet[][orange solid line]{2018A&A...615A..27L}, the \citet[][green dots]{2018MNRAS.475.2891D}, and the \citet[][purple dots]{2019MNRAS.482.1557S} compilations of literature estimates. As with SFRD($z$), a trend is similar to those reported on previous studies based mostly on large cosmological surveys. SMD grows rapidly in the early cosmological epochs, increasing from $\sim 10^{6.5} \mathrm{M_{\odot} \; Mpc}^{-3} $ in the first 3-4 Gyrs, reaching a level above $\sim 10^{8.2} \mathrm{M_{\odot} \; Mpc}^{-3}$ by the current era. This shape of the SMD curve has been well described in works based on cosmological surveys \citep[][and others]{2014ARA&A..52..415M, 2018A&A...615A..27L}. The obtained result is slightly lower than the values from previous studies, which is explained by the difference in the operation of the SED-fitting algorithms, as well as the already mentioned cosmic variance.

\section{Conclusions}

We have presented gMOSS, a new medium-band photometric catalogue that covers 2.386 $\mathrm{deg^2}$ of the HS47.5-22 field, which includes measurements of 16 medium-band and 4 broadband filters of the SDSS system (we publish general fluxes and corresponding uncertainties of $1\sigma$), and photometric redshifts of 19,875 galaxies. The sample of galaxies in the gMOSS survey is limited by threshold magnitude $r_{\mathrm{AB}}=22.5$ mag. Also, we present the spectral redshifts obtained from observations at the 6-meter Russian telescope with SCORPIO-2 multi-mode focus reducer for selected galaxies in the HS47.5-22 field.

We applied the SED-fitting code \textsc{cigale} to extract the star formation history of 12,281 galaxies up to $z = 0.8$. We used a set of spectral evolution synthesis models from the BC03 set \cite{2003MNRAS.344.1000B}, with the modified attenuation law by \cite{2000ApJ...533..682C} and assuming the initial mass function of \cite{2003PASP..115..763C}. We have obtained the stellar mass, rest frame and internal $(u-r)$ colours, extinction, age and metallicity of galaxy stellar populations. The standard deviations in the stellar population parameters are: $0.19 \pm 0.08$ mag, $0.12 \pm 0.05$ dex, $0.10 \pm 0.04$ dex, and $0.28 \pm 0.09$ mag for $(u - r)_{\mathrm{res}}$ colour, total stellar mass, mass-weighted age, and extinction, respectively. For the metallicity parameter, we used a set of 6 discrete values, so this parameter is less amenable to analysis.

We have divided the galaxies into blue and red populations. The easiest way to split a full galaxy sample is to use the dust-corrected $(u - r)_{\mathrm{int}}$ diagram. The percentage of the number of the red sequence galaxies and the blue cloud galaxies is 14 per cent and 86 per cent, respectively. The properties of stellar populations clearly differ between red sequence and blue cloud galaxies. Red and blue galaxies are identified by their $(u - r)_{\mathrm{int}}$ colour, with mean values $\sim 2$ and $\sim 1$. Throughout the full redshift range ($0 < z \le 0.8$), red galaxies are older and more massive than blue ones.

The evolution of galaxies with redshift shows that at the present cosmic time galaxies are older than at $z = 0.8$. The average mass of galaxies increases with increasing redshift, which is a manifestation of the selection of a flux-limited sample.

As part of the study of the star formation rate density, we determined our sample is complete for galaxies with above $\mathrm{log}(M)\!_{[\mathrm{M_\odot}]} \sim 8.9$, 9.5, and 9.9 dex at $z = 0.4$, 0.6, and 0.8, respectively. Based on the obtained SFRD data at $z \sim 0.1$, we can estimate the evolution of SFRD and SMD, getting a result that corresponds to the results of previously published works of cosmological surveys. This adds to the weight of evidence that the CSFH is now well known.

\section*{Acknowledgements}
The work was carried out within the framework of the government contract of SAO RAS approved by the Ministry of Science and Higher Education of the Russian Federation. Observations with the SAO RAS telescopes are supported by the Ministry of Science and Higher Education of the Russian Federation (including agreement No05.619.21.0016, project ID RFMEFI61919X0016).

This research has made use of the NASA/IPAC Extragalactic Database (NED), which is operated by the Jet Propulsion Laboratory, California Institute of Technology, under contract with the National Aeronautics and Space Administration.

Funding for the Sloan Digital Sky Survey IV has been provided by the Alfred P. Sloan Foundation, the U.S. Department of Energy Office of Science, and the Participating Institutions. SDSS-IV acknowledges support and resources from the Center for High-Performance Computing at the University of Utah. The SDSS web site is \url{www.sdss.org}.

This work has made use of data from the European Space Agency (ESA) mission
{\it Gaia} (\url{https://www.cosmos.esa.int/gaia}), processed by the {\it Gaia}
Data Processing and Analysis Consortium (DPAC,
\url{https://www.cosmos.esa.int/web/gaia/dpac/consortium}). Funding for the DPAC
has been provided by national institutions, in particular the institutions
participating in the {\it Gaia} Multilateral Agreement.

The Legacy Surveys consist of three individual and complementary projects: the Dark Energy Camera Legacy Survey (DECaLS; Proposal ID \#2014B-0404; PIs: David Schlegel and Arjun Dey), the Beijing-Arizona Sky Survey (BASS; NOAO Prop. ID \#2015A-0801; PIs: Zhou Xu and Xiaohui Fan), and the Mayall z-band Legacy Survey (MzLS; Prop. ID \#2016A-0453; PI: Arjun Dey). DECaLS, BASS and MzLS together include data obtained, respectively, at the Blanco telescope, Cerro Tololo Inter-American Observatory, NSF’s NOIRLab; the Bok telescope, Steward Observatory, University of Arizona; and the Mayall telescope, Kitt Peak National Observatory, NOIRLab. The Legacy Surveys project is honored to be permitted to conduct astronomical research on Iolkam Du’ag (Kitt Peak), a mountain with particular significance to the Tohono O’odham Nation.

NOIRLab is operated by the Association of Universities for Research in Astronomy (AURA) under a cooperative agreement with the National Science Foundation.

This project used data obtained with the Dark Energy Camera (DECam), which was constructed by the Dark Energy Survey (DES) collaboration. Funding for the DES Projects has been provided by the U.S. Department of Energy, the U.S. National Science Foundation, the Ministry of Science and Education of Spain, the Science and Technology Facilities Council of the United Kingdom, the Higher Education Funding Council for England, the National Center for Supercomputing Applications at the University of Illinois at Urbana-Champaign, the Kavli Institute of Cosmological Physics at the University of Chicago, Center for Cosmology and Astro-Particle Physics at the Ohio State University, the Mitchell Institute for Fundamental Physics and Astronomy at Texas A\& M University, Financiadora de Estudos e Projetos, Fundacao Carlos Chagas Filho de Amparo, Financiadora de Estudos e Projetos, Fundacao Carlos Chagas Filho de Amparo a Pesquisa do Estado do Rio de Janeiro, Conselho Nacional de Desenvolvimento Cientifico e Tecnologico and the Ministerio da Ciencia, Tecnologia e Inovacao, the Deutsche Forschungsgemeinschaft and the Collaborating Institutions in the Dark Energy Survey. The Collaborating Institutions are Argonne National Laboratory, the University of California at Santa Cruz, the University of Cambridge, Centro de Investigaciones Energeticas, Medioambientales y Tecnologicas-Madrid, the University of Chicago, University College London, the DES-Brazil Consortium, the University of Edinburgh, the Eidgenossische Technische Hochschule (ETH) Zurich, Fermi National Accelerator Laboratory, the University of Illinois at Urbana-Champaign, the Institut de Ciencies de l’Espai (IEEC/CSIC), the Institut de Fisica d’Altes Energies, Lawrence Berkeley National Laboratory, the Ludwig Maximilians Universitat Munchen and the associated Excellence Cluster Universe, the University of Michigan, NSF’s NOIRLab, the University of Nottingham, the Ohio State University, the University of Pennsylvania, the University of Portsmouth, SLAC National Accelerator Laboratory, Stanford University, the University of Sussex, and Texas A\&M University.

BASS is a key project of the Telescope Access Program (TAP), which has been funded by the National Astronomical Observatories of China, the Chinese Academy of Sciences (the Strategic Priority Research Program “The Emergence of Cosmological Structures” Grant \# XDB09000000), and the Special Fund for Astronomy from the Ministry of Finance. The BASS is also supported by the External Cooperation Program of Chinese Academy of Sciences (Grant \# 114A11KYSB20160057), and Chinese National Natural Science Foundation (Grant \# 11433005).

The Legacy Survey team makes use of data products from the Near-Earth Object Wide-field Infrared Survey Explorer (NEOWISE), which is a project of the Jet Propulsion Laboratory/California Institute of Technology. NEOWISE is funded by the National Aeronautics and Space Administration.

The Legacy Surveys imaging of the DESI footprint is supported by the Director, Office of Science, Office of High Energy Physics of the U.S. Department of Energy under Contract No. DE-AC02-05CH1123, by the National Energy Research Scientific Computing Center, a DOE Office of Science User Facility under the same contract; and by the U.S. National Science Foundation, Division of Astronomical Sciences under Contract No. AST-0950945 to NOAO.

\section*{Data Availability}

All data is available to the reviewer(s)and will be made open-source on publication. 



\bibliographystyle{mnras}
\bibliography{example} 

\bsp	
\label{lastpage}
\end{document}